\documentclass[twocolumn,secnumarabic,amssymb, nobibnotes, aps, prx]{revtex4-2}

\usepackage{graphicx}
\usepackage{float}
\usepackage{longtable}
\usepackage{mwe}
\usepackage{dsfont}
\usepackage{bbm}
\usepackage{xcolor}
\usepackage{amsmath}
\usepackage{verbatim}
\usepackage{cleveref}
\newcommand*{\colorboxed}{}
\def\colorboxed#1#{%
  \colorboxedAux{#1}%
}
\newcommand*{\colorboxedAux}[3]{%
  \begingroup
    \colorlet{cb@saved}{.}%
    \color#1{#2}%
    \boxed{%
      \color{cb@saved}%
      #3%
    }%
  \endgroup
}
\begin{document}

\title{Collective dynamics Using Truncated Equations (CUT-E): simulating the collective strong coupling regime with few-molecule models}%

\author{Juan B. P\'erez-S\'anchez, Arghadip Koner, Joel Yuen-Zhou}%
\affiliation{Department of Chemistry and Biochemistry, University of California San Diego, La Jolla, California, 92093, United States}
\email[Email: ]{joelyuen@ucsd.edu}
\author{Nathaniel P. Stern}
\affiliation{Department of Physics and Astronomy, Northwestern University, Evanston, Illinois, 60208, United States}
\date{\today}%

\begin{abstract}
The study of molecular polaritons beyond simple quantum emitter ensemble models (e.g., Tavis-Cummings) is challenging due to the large dimensionality of these systems (the number of molecular emitters is $N\approx 10^{6}-10^{10}$) and the complex interplay of molecular electronic and nuclear degrees of freedom. This complexity constraints existing models to either coarse-grain the rich physics and chemistry of the molecular degrees of freedom or artificially limit the description to a small number of molecules. In this work, we exploit permutational symmetries to drastically reduce the computational cost of \textit{ab-initio} quantum dynamics simulations for large $N$. Furthermore, we discover an emergent hierarchy of timescales present in these systems, that justifies the use of an \textit{effective} single molecule to approximately capture the dynamics of the entire ensemble, an approximation that becomes exact as $N\rightarrow \infty$. We also systematically derive finite $N$ corrections to the dynamics, and show that addition of $k$ extra effective molecules is enough to account for phenomena whose rates scale as $\mathcal{O}(N^{-k})$.  Based on this result, we discuss how to seamlessly modify existing single-molecule strong coupling models to describe the dynamics of the corresponding ensemble, as well as the crucial differences in phenomena predicted by each model. We call this approach Collective dynamics Using Truncated Equations (CUT-E), benchmark it against well-known results of polariton relaxation rates, and apply it to describe a universal cavity-assisted energy funneling mechanism between different molecular species. Beyond being a computationally efficient tool, this formalism provides an intuitive picture for understanding the role of bright and dark states in chemical reactivity, necessary to generate robust strategies for polariton chemistry.
\end{abstract}

\maketitle

\section{Introduction}

Molecular polaritons are quasiparticles arising when vibrational and/or electronic excitations of an ensemble of molecules are collectively coupled to a confined electromagnetic mode such as those found in optical microcavities. These  systems have attracted much interest in the last decade due to their potential applications in altering chemical reactions dynamics \cite{Ebbesen,Herrera,Sindhana,RaphaelReview,Huo,Neepa,Groenhof,Feist1}, energy transfer and energy conversion mechanisms \cite{Coles,Zhong,Reitz,DelPo,Genes,Ebbesen2,Wei,Luis,SKC1,SKC2}, and as a framework to achieve room-temperature exciton-polariton condensation \cite{SKC3,SKC4}. Theoretical work aimed at explaining experimental results or predicting new phenomena emerging in polaritonic architectures face the formidable challenge of properly modeling the molecular (local) degrees of freedom of each molecule while describing the super-radiant interaction of the molecular ensemble with the field (collective). The dynamics arising from the complex interplay of vibrational and electronic degrees of freedom in molecules renders simple quantum optics models (such as the original Tavis-Cummings Hamiltonian \cite{TC}), limited in their applicability to molecular polaritons. Thus, molecular polaritons face unique challenges and opportunities that are not encountered in more traditional polariton systems \cite{Feist5}, such as atomic or artificial qubit ensembles \cite{qubits}, or cryogenic inorganic semiconductors \cite{cryo}. Most of the reported simulations of molecular polaritons can only deal with one of two aforementioned challenges at a time. On the one hand, theoretical studies that acknowledge the collective nature of the light-matter coupling are typically limited to a few dozen molecules at a time and involve sophisticated numerical treatments \cite{delPino2}, simplifications such as single vibrational mode descriptions \cite{HerreraPRL2}, or semiclassical trajectories \cite{Groenhof,Groenhof3}. On the other hand, models that implement \textit{ab initio} treatments are often restricted to a single or few molecules in a cavity \cite{Dominik1,Dominik2,Schafer}. Regardless, from a computational standpoint, it seems suspicious that it is necessary to explicitly simulate the dynamics of $N$ molecules, especially if they are identical to each other. Indeed, there are numerous symmetries in the system that should significantly reduce the computational cost of these simulations \cite{Keeling,Spano,Spano2,Spano3,NoriPerm,Herrera2,Silva,KeelingZeb2,Jorge,Cederbaum,Richter2}. 

In this work, we outline a wavefunction-based formalism that makes use of such symmetries to significantly reduce the complexity of quantum dynamics simulations of the single-excitation manifold of molecular polaritons. Moreover, this formalism naturally provides a hierarchy of approximations to further simplify the problem in a way that, in the $N\rightarrow \infty$ limit, polaritonic properties can be calculated using a modified \textit{effective} single molecule coupled to a cavity with the collective coupling. This provides grounds for some single-molecule strong coupling phenomena to appear in the collective regime, consistent with previous work where linear optical properties can be calculated from effective single-molecule models in the thermodynamic limit \cite{KeelingZeb,Cwik}. Moreover, we show that a system of two effective molecules can describe all the effects with rates that scale as $1/N$. In general, we show that processes with $\mathcal{O}(N^{-k})$ rates are described by $k+1$ effective molecules strongly coupled to the cavity. This implies that, for a large ensemble of molecules, it is enough to consider only a few effective molecules to solve for the dynamics of the original polariton system (see Fig. \ref{micro}). The model can be applied to study disordered ensembles (e.g., a mixture with two chemical species) without a significant increase in the computational cost. 

The article is organized as follows: in Section \ref{sec:theory} we present the Hamiltonian and the multiconfigurational representation of the total wavefunction of the system, in which permutational symmetries become evident. Then, we uncover a convenient mathematical structure of the Equations of Motion (EoM) where approximate symmetries emerge, and which become useful for large $N$, while keeping the collective light-matter coupling $\sqrt{N}g$ finite, with $g$ being the single-molecule coupling (this is the physical condition of interest in experiments, and which concerns us hereafter). This structure allows us to derive the simple effective Hamiltonians involving only a few molecules, that solve for the dynamics of the entire ensemble. In Section \ref{sec:zero} we make use of the effective single-molecule model to demonstrate how both optical and material properties in the original system can be computed using the effective single-molecule simulation. In Section \ref{sec:relax} we benchmark our formalism against a well known result: the non-radiative relaxation of polariton and dark states. In Section \ref{sec:nonstat} we present a pedagogical and intriguing application that reveals the power of our formalism, describing how to exploit polariton dynamics to obtain nonstatistical outcomes in photoproducts. Finally we summarize the work in Section \ref{sec:summ}.

\section{Theory}\label{sec:theory}

\subsection{Hamiltonian and Multi-Configurational Wavefunction}

\begin{figure}[]
\begin{center}
\includegraphics[width=1\linewidth]{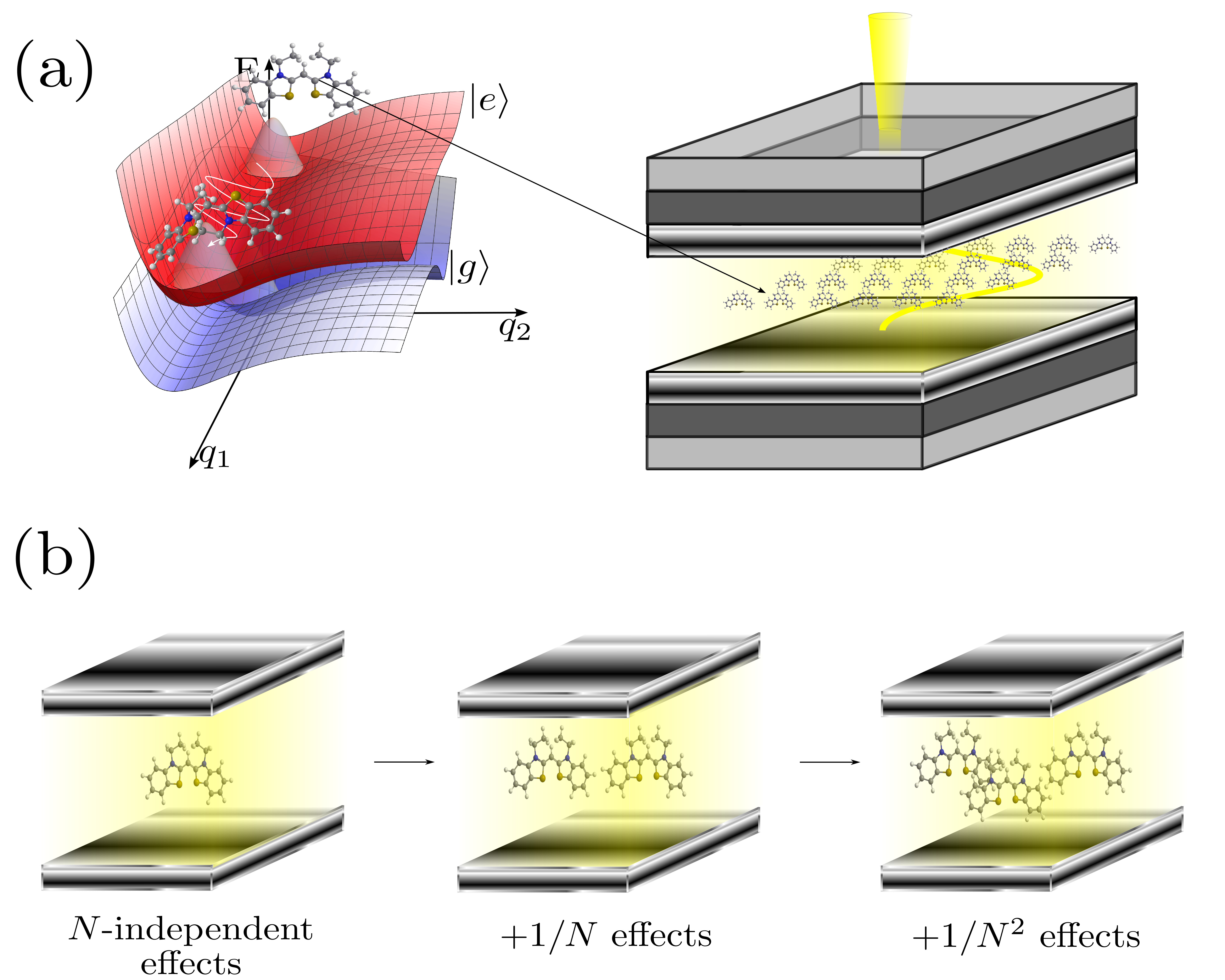}
\caption{a) Molecular polaritons in the collective strong-coupling regime. Molecules are not well described by structureless two-level systems, and the interplay between their internal (e.g., vibrational) degrees of freedom and the collective interaction of their optical (e.g., electronic) transitions with the optical mode  cannot be described using simple models such as the standard Tavis-Cummings Hamiltonian. In fact, the modified internal dynamics of the molecules under collective strong coupling is the subject of study of polariton chemistry. b) Pictorial representation of the Collective dynamics Using Truncated Equations (CUT-E) method. Processes with $\mathcal{O}(N^{-k})$ rates can be described by a model of $k+1$ effective molecules coupled to a cavity mode.}
\label{micro}
\end{center}
\end{figure}

Consider a system of $N$ molecules collectively coupled to a single cavity mode. The Tavis-Cummings Hamiltonian, extended to include vibrational degrees of freedom missing from original models, can be written as (hereafter $\hbar=1$) 

\begin{equation}\label{eq:ham}
\hat{H}=\sum_{i}^{N}\left(\hat{H}_{m}^{(i)}+\hat{H}^{(i)}_{I}\right)+\hat{H}_{cav},
\end{equation}where 

\begin{align*}
&\hat{H}^{(i)}_{m}=-\frac{1}{2\mu}\frac{\partial^{2}}{\partial q_{i}^{2}}+V_{g}(q_{i})|g_{i}\rangle\langle g_{i}|+V_{e}(q_{i})|e_{i}\rangle\langle e_{i}|,\\
&\hat{H}_{cav}=\omega_{c}\hat{a}^{\dagger}\hat{a},\ \ \ \ \ \ \ \hat{H}^{(i)}_{I}=g\left(|e_{i}\rangle\langle g_{i}|\hat{a}+|g_{i}\rangle\langle e_{i}|\hat{a}^{\dagger}\right),
\end{align*} are the Hamiltonians for the $i$th molecule, the cavity mode, and the interaction between them. Here, $\mu$ is the reduced mass of the nuclei, $|g_{i}\rangle$ and $|e_{i}\rangle$ are the molecular ground and excited electronic states, $V_{g/e}(q_{i})$ are the ground and excited Potential Energy Surfaces (PES), $\hat{a}$ is the photon annihilation operator, and $q_{i}$ is the vector of all molecular vibrations of molecule $i$ (see Fig. \ref{micro}). When the PESs are harmonic, the model above reduces to the Holstein-Tavis Cummings model, which has been subject of recent studies \cite{HerreraPRL2,KeelingZeb}. For the time being, we assume there is no disorder, neglect intermolecular interactions, and use the rotating wave approximation by considering the single-molecule coupling strength $g$ to be much smaller than the bare photon frequency $\omega_{c}$. Finally, we also ignore (non-radiative) couplings between the molecular ground and excited states whose PESs can form conical intersections \cite{Vibok,VendrellPRL,Csehi}. Under these hypotheses, the excitation number [sum of electronic excitations (Frenkel excitons) and photon number] is conserved. 

In this article we shall focus on the so-called first-excitation manifold, which affords a position-representation ansatz of the form,

\begin{equation}\label{eq:exact1}
|\Psi(t)\rangle=\psi^{(0)}(\vec{q},t)|1\rangle+\sum_{i}^{N}\psi^{(i)}(\vec{q},t)|e_{i}\rangle,
\end{equation}with the electron-photon states $|e_{i}\rangle=|g_{1},g_{2},...,e_{i},...,g_{N},0_{ph}\rangle$ and $|1\rangle=|g_{1},g_{2},...,g_{N},1_{ph}\rangle$, and the vibrational wavefunctions

\begin{equation}\label{eq:exact2}
\psi^{(0)}(\vec{q},t)=\sum_{j_{1}}^{m}\sum_{j_{2}}^{m}\cdots\sum_{j_{N}}^{m}A^{(0)}_{j_{1}j_{2}...j_{N}}(t)\prod_{k=1}^{N}\varphi_{j_{k}}(q_{k}),
\end{equation} and

\begin{equation}\label{eq:exact3}
\psi^{(i)}(\vec{q},t)=\sum_{j_{1}}^{m}\sum_{j_{2}}^{m}\cdots\sum_{j_{N}}^{m}A^{(i)}_{j_{1}j_{2}...j_{N}}(t)\phi_{j_{i}}(q_{i})\prod_{k\neq i}^{N}\varphi_{j_{k}}(q_{k}).
\end{equation} In this expansion, multiconfigurational vibrational wavefunctions $\psi$ are built with sets of $m$ single particle orthonormal functions $\varphi_{j_{i}}(q_{i})$ and $\phi_{j_{i}}(q_{i})$, respectively, that are equal for identical molecules. In this work, we choose such functions to be the eigenstates of the molecular Hamiltonian $\hat{H}^{(i)}_{m}$. The total wavefunction becomes exact as $m\rightarrow \infty$. 

\subsection{Exploiting Permutational Symmetries}

Since the Hamiltonian is invariant under the permutation of any pair of molecules $\alpha$ and $\kappa$, having an initial state such that $P_{\alpha\kappa}|\Psi(0)\rangle=|\Psi(0)\rangle$, implies permutation relations between the coefficients of the wavefunction. For the photonic and excitonic wavefunctions $\psi^{(0)}$ and $\psi^{(i)}$ we have, 

\begin{align}\label{eq:perm1}
&A^{(0)}_{j_{1}j_{2}...j_{\kappa}...j_{\nu}...j_{N}}(t)= A^{(0)}_{j_{1}j_{2}...j_{\nu}...j_{\kappa}...j_{N}}(t),\nonumber\\
&A^{(i)}_{j_{1}j_{2}...j_{\kappa}...j_{i}...j_{\nu}...j_{N}}(t)= A^{(i)}_{j_{1}j_{2}...j_{\nu}...j_{i}...j_{\kappa}...j_{N}}(t).
\end{align} Additionally, there are permutations between coefficients of different excitons, given the interaction of the molecules with the cavity is assumed identical,

\begin{equation}\label{eq:perm2}
A^{(i)}_{j_{1}j_{2}...j_{i}...j_{i'}...j_{N}}(t)= A^{(i')}_{j_{1}j_{2}...j_{i'}...j_{i}...j_{N}}(t).
\end{equation} This means that every coefficient of the electronic state $i'$ can be obtained directly from those in the electronic state $i$. In other words, Eqs. \ref{eq:perm1} and \ref{eq:perm2} yield the crucial observation that it is enough to calculate the dynamics of a single excitonic state to know the evolution of all of them. Moreover, the vibrational states of the ensemble of molecules can be completely characterized by specifying the number of molecules in each vibrational state; therefore the number of vibrational degrees of freedom also reduces drastically. The ground and excited state coefficients can be written in terms of permutationally-symmetric states,
\begin{align}\label{eq:notation}
&A^{(0)}_{j_{1}j_{2}j_{3}\cdots j_{N}}\rightarrow A^{(0)}_{N_{1}N_{2}\cdots N_{m}}\nonumber\\
&A^{(i)}_{j_{1}j_{2}j_{3}\cdots j_{N}}\rightarrow A^{(1)}_{j_{1}N_{1}N_{2}\cdots N_{m}},
\end{align}where $N_{k}$ is the number of ground state molecules in the vibrational state $k$.  This new notation removes the information about the state of each specific molecule ($A^{(1)}_{j_{1}N_{1}N_{2}\cdots N_{m}}$ represents a state where one molecule is in the electronic excited state in the vibrational state $j_{1}$, while the other $N-1$ molecules are distributed among all the vibrational states via \{$N_{k}$\}, $\sum_{k}N_{k}=N-1$. For the photonic coefficients $A^{(0)}_{N_{1}N_{2}\cdots N_{m}}$, the restriction is $\sum_{k}N_{k}=N$. For the vibrational wavefunction $\psi^{(0)}$ not all $m^{N}$ configurations are unique but only $\binom{N+m-1}{N}$. Using similar analysis we conclude a reduction of the total wavefunction in Eq. \ref{eq:exact1} from $(N+1)m^{N}$ to $\binom{N+m-1}{N}+m\binom{N+m-2}{N-1}$ configurations. This corresponds to reducing the complexity from an exponential scaling to a polynomic one, as has been shown in previous work that also exploit permutational symmetries for studying ensembles of identical systems \cite{Richter2,Richter}. We will make use of these simplifications to redefine the EoM and subsequently the Hamiltonians. Such analysis is general and can also be done for higher excitation manifolds. In that case, the number of coefficients increases with the number of excitations due to permutations between electronically excited molecules; however after the number of excitations is half of the number of molecules this trend reverses.

The EoM for the coefficients can be obtained by using the Dirac-Frenkel variational principle, or simply by inserting the ansatz wavefuntion into the Time-Dependent Schr\"odinger Equation. Rewriting the ansatz as $|\Psi(t)\rangle=\sum_{J}^{m^{N}}A^{(0)}_{J}(t)\Phi^{(0)}_{J}|1\rangle+\sum_{i}\sum_{J}^{m^{N}}A^{(i)}_{J}(t)\Phi^{(i)}_{J}|e_{i}\rangle$ the EoM become \cite{mctdh}

\begin{equation}\label{eq:EoM1}
i\dot{A}^{(i)}_{J}(t)=\sum_{i'}\sum_{L}\langle \Phi^{(i)}_{J}|\hat{H}|\Phi^{(i')}_{L}\rangle A^{(i')}_{L}(t).
\end{equation} Since Eq. \ref{eq:ham} ignores couplings between the molecules in the absence of the photon mode, $\langle \Phi^{(i)}_{J}|\hat{H}_{I}|\Phi^{(i')}_{L}\rangle= 0$ for $i,i'\neq 0$. Furthermore,

\begin{equation}\label{eq:EoM2}
\langle \Phi^{(i)}_{J}|\hat{H}_{I}|\Phi^{(0)}_{L}\rangle=g\langle \phi_{j_{i}}|\varphi_{l_{i}}\rangle\prod_{k\neq i}^{N}\delta_{j_{k}l_{k}}.
\end{equation}With these considerations, Eq. \ref{eq:EoM1} becomes,

\begin{align}\label{eq:EoM3}
i\dot{A}^{(i)}_{j_{1}j_{2}...j_{i}...j_{N}}(t)&=\left(\sum_{i'\neq i}^{N}E_{g,j_{i'}}+E_{e,j_{i}}\right)A^{(i)}_{j_{1}j_{2}...j_{i}...j_{N}}(t)\nonumber\\
&+g\sum_{l_{i}}\langle \phi_{j_{i}}|\varphi_{l_{i}}\rangle A^{(0)}_{j_{1}j_{2}...l_{i}...j_{N}}(t)\nonumber\\
i\dot{A}^{(0)}_{j_{1}j_{2}...j_{N}}(t)&=\left(\sum_{i'}^{N}E_{g,j_{i'}}+\omega_{c}\right)A^{(0)}_{j_{1}j_{2}...j_{N}}(t)\nonumber\\
&+g\sum_{i=1}^{N}\sum_{l_{i}}\langle \varphi_{j_{i}}|\phi_{l_{i}}\rangle A^{(i)}_{j_{1}j_{2}...l_{i}...j_{N}}(t).
\end{align} The last equation can be simplified using Eqs. \ref{eq:notation} to write the dynamics in terms of permutationally-symmetric states:

\begin{align}\label{eq:EoM4}
&i\dot{A}^{(1)}_{lN_{1}N_{2}\cdots N_{m}}(t)=\left(\sum_{k=1}^{m}N_{k}E_{g,k}+E_{e,l}\right)A^{(1)}_{lN_{1}N_{2}\cdots N_{m}}(t)\nonumber\\
&+g\sum_{k}\langle \phi_{l}| \varphi_{k}\rangle A^{(0)}_{N_{1}N_{2}\cdots N_{k}+1 \cdots N_{m}}(t)\nonumber\\
&i\dot{A}^{(0)}_{N_{1}N_{2}\cdots N_{m}}(t)=\left(\sum_{k=1}^{m}N_{k}E_{g,k}+\omega_{c}\right)A^{(0)}_{N_{1}N_{2}\cdots N_{m}}(t)\nonumber\\
&+\sum_{k=1}^{m}N_{k}g\sum_{l=1}^{m}\langle \varphi_{k}| \phi_{l}\rangle A^{(1)}_{lN_{1}\cdots N_{k}-1 \cdots N_{m}}(t).
\end{align} The first equation represents the absorption of the photon that takes a molecule from the state $\varphi_{k}|g\rangle$ into the state $\phi_{l}|e\rangle$. The second equation describes the conjugate process. 

\subsection{Structure of the Wavefunction}

\begin{figure}[]
\begin{center}
\includegraphics[width=1\linewidth]{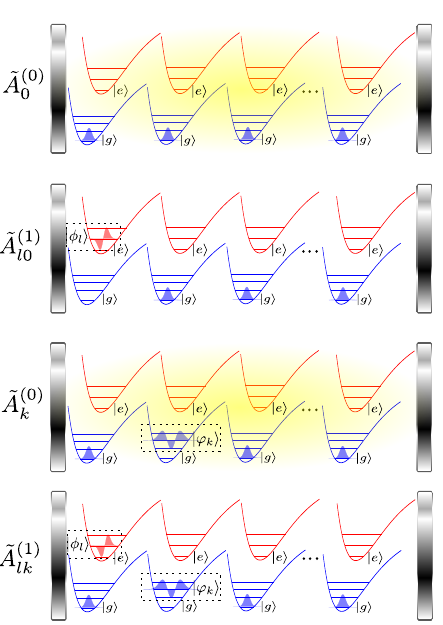}
\caption{Permutationally symmetric basis in the first excitation manifold, originally introduced by Spano [\citenum{Spano}]. $\tilde{A}^{(0)}_{0}:$ a photon in the cavity and $N$ molecules in the global ground state. $\tilde{A}^{(1)}_{l0}:$ $1$ exciton in the vibrational state $l$ and $N-1$ molecules are in the global ground state. $\tilde{A}^{(0)}_{k}:$ a photon in the cavity, a ground state molecule in the vibrational state $k$, and $N-1$ molecules in the global ground state. $\tilde{A}^{(1)}_{lk}:$ $1$ exciton in the vibrational state $l$, $1$ ground state molecule in the vibrational state $k$, and $N-1$ molecules in the global ground state.}
\label{spanos}
\end{center}
\end{figure}

Even though Eqs. \ref{eq:EoM4} are exact and represent a significant improvement over Eqs. \ref{eq:EoM3}, they also allow us to systematically introduce approximations by virtue of the factors $N_{k}$, which represent the number of ground state molecules in the vibrational state $k$. For initial states in which one of the values $N_{k}$ is exceptionally large, the dynamics is such that $N_{k}$ is almost conserved, as discussed below. As an example, assume we start with all the molecules in their ground vibrational state and $1$ photon in the cavity mode, i.e., $A^{(0)}_{N00\cdots 0}(0)=1$. Thus, we can simplify even more our notation by only reporting the number of such vibrational excitations. By renormalizing the amplitudes, we can recover the basis originally introduced by Spano for harmonic modes, but which we now use for arbitrary PESs [\citenum{Spano}]:

\begin{align}\label{eq:renorm}
&\tilde{A}^{(0)}_{0}(t)=A^{(0)}_{N00\cdots 0}(t),\nonumber\\
&\tilde{A}^{(1)}_{l0}(t)=\sqrt{N}A^{(1)}_{l(N-1)0\cdots 0}(t),\nonumber\\
&\tilde{A}^{(0)}_{k}(t)=\sqrt{N}A^{(0)}_{(N-1)\cdots 1_{k}\cdots 0}(t),\nonumber\\
&\tilde{A}^{(1)}_{lk}(t)=\sqrt{N(N-1)}A^{(1)}_{l(N-2)\cdots 1_{k}\cdots 0}(t),\nonumber\\
&\tilde{A}^{(0)}_{k\neq k'}(t)=\sqrt{N(N-1)}A^{(0)}_{(N-2)\cdots 1_{k}\cdots 1_{k'}\cdots 0}(t),\nonumber\\
&\tilde{A}^{(0)}_{kk}(t)=\sqrt{\frac{N(N-1)}{2}}A^{(0)}_{(N-2)\cdots 2_{k}\cdots 0}(t), 
\end{align}and so on. With this final notation, the EoM in Eq. \ref{eq:EoM4} become,

\begin{widetext}
\begin{align}\label{eq:EoM5}
&i\dot{\tilde{A}}^{(0)}_{0}(t)=\omega_{c}\tilde{A}^{(0)}_{0}(t)+g\sqrt{N}\sum_{l=1}^{m}\langle \varphi_{1}|\phi_{l}\rangle \tilde{A}^{(1)}_{l0}(t)\nonumber\\
&i\dot{\tilde{A}}^{(1)}_{l0}(t)=\omega_{eg,l}\tilde{A}^{(1)}_{l0}(t)+g\sqrt{N}\langle \phi_{l}|\varphi_{1}\rangle \tilde{A}^{(0)}_{0}(t)+g\sum_{k=2}^{m}\langle \phi_{l}|\varphi_{k}\rangle \tilde{A}^{(0)}_{k}(t)\nonumber\\
&i\dot{\tilde{A}}^{(0)}_{k}(t)=\left(\omega_{g,k}+\omega_{c}\right)\tilde{A}^{(0)}_{k}(t)+g\sqrt{N-1}\sum_{l=1}^{m}\langle \varphi_{1}|\phi_{l}\rangle \tilde{A}^{(1)}_{lk}(t)+g\sum_{l=1}^{m}\langle \varphi_{k}|\phi_{l}\rangle \tilde{A}^{(1)}_{l0}(t)\nonumber\\
&i\dot{\tilde{A}}^{(1)}_{lk}(t)=\left(\omega_{eg,l}+\omega_{g,k}\right)\tilde{A}^{(1)}_{lk}(t)+g\sqrt{N-1}\langle \phi_{l}|\varphi_{1}\rangle A^{(0)}_{k}(t)+g\sum_{k'\neq k}\langle \phi_{l}|\varphi_{k'}\rangle \tilde{A}^{(0)}_{kk'}(t)+\sqrt{2}g\langle \phi_{l}|\varphi_{k}\rangle \tilde{A}^{(0)}_{kk}(t)\nonumber\\
& \hspace{3in}\vdots
\end{align}
\end{widetext}where we have removed a constant $NE_{g,1}$ and defined $\omega_{eg,l}=E_{e,l}-E_{g,1}$ and $\omega_{g,k}=E_{g,k}-E_{g,1}$. In Fig. \ref{spanos} we provide a pictorial representation of the states associated with these coefficients.

The first equation reveals that the initial photonic state $\tilde{A}^{(0)}_{0}(t)$ is strongly coupled to states in which one molecule is electronically excited while the rest remain in their ground electronic and vibrational states $\tilde{A}^{(1)}_{l0}(t)$. However, the second equation reveals that such excited state can either emit back into the initial state (with no phonons), or can create states in which there is a vibrational excitation in one of the ground state molecules $\tilde{A}^{(0)}_{k}(t)$ upon emission. The first of these two processes depends on the collective light-matter coupling $\sqrt{N}g$, while the second one depends on the single-molecule light-matter coupling $g$. This structure is repeated throughout the system of equations: coupling between states conserving the number of molecules with phonons is collective, while processes that increase the number of such molecules are proportional to the single-molecule coupling. It is interesting to note that this phenomenon is well-known in the literature of molecular aggregates. In particular, for J-aggregates, Spano and Yamagata \cite{Spano3} have noted that the ratio of the photoluminescence into the electronic ground state with no phonons versus that into the electronic ground state with one phonon is proportional to the coherence length $N$ of the aggregate. While this phenomenon is routinely used as a spectroscopic probe for $N$, we use it in our case to drastically simplify the simulations of molecular polaritons, as explained below.    

Assuming we have a large number of molecules we can treat the mixing of states with differing number of ground state molecules with phonons perturbatively, with the \textit{single}-molecule light-matter coupling $g$ as the perturbation. This is schematically represented in the next figure.

\begin{figure*}[htp]
\begin{center}
\includegraphics[width=1\linewidth]{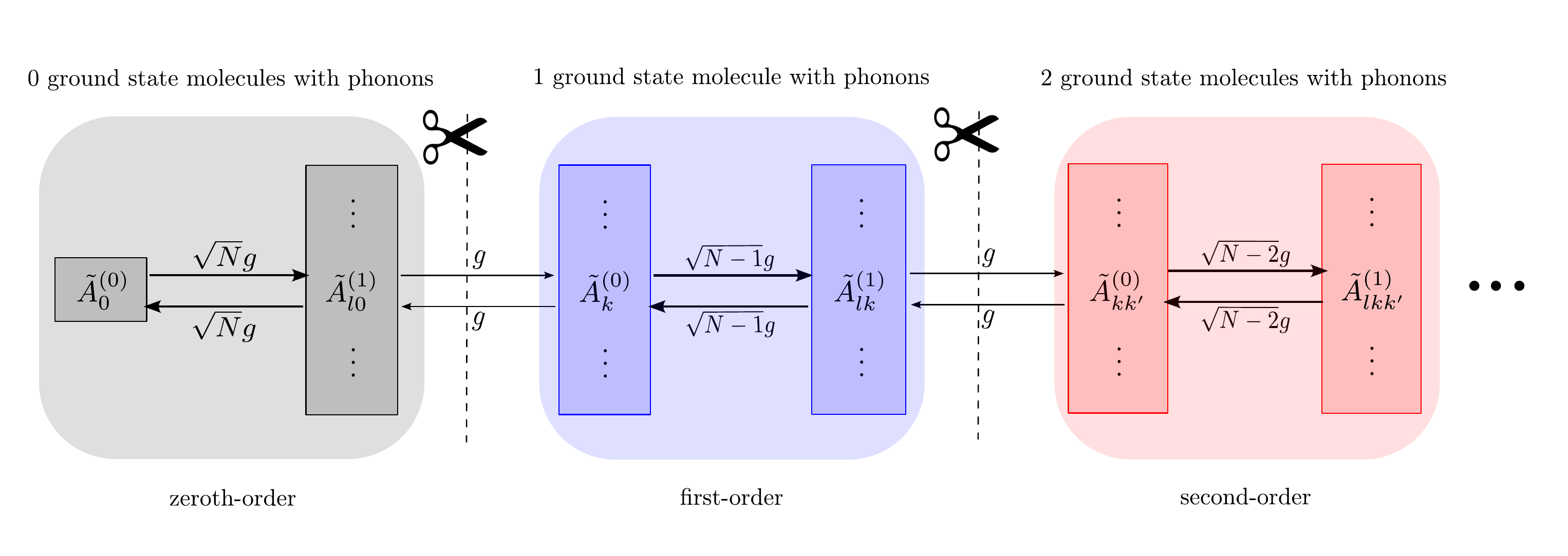}
\caption{Hierarchical structure of the Equations of Motion (EoM) that gives rise to the Collective dynamics Using Truncated Equations (CUT-E) method. Notice that fast dynamics in each order of approximation conserves the number of electronic ground state molecules with phonons (due to collective $\sqrt{N}g$ couplings). These fast dynamics are linked by bottlenecks (due to single-molecule $g$ couplings) which slowly change the number of molecules featuring such ground-state phonons. Zeroth-order approximation in $g$ corresponds to restricting the dynamics to states with a fixed number of ground state molecules with phonons, while adding the first-order correction allows the dynamics to create (or annihilate) phonons in (from) $1$ additional molecule.}
\label{wf_dyn2}
\end{center}
\end{figure*}

In the limit where $g\rightarrow 0$ or $N\rightarrow \infty$ (while keeping $\sqrt{N}g$ constant), which is the regime that interests us, the number of ground state molecules that support vibrational excitations is conserved during the dynamics. For our initial state, the zeroth-order approximation implies that the wavefunction is described only by the basis states of the left most block of Fig. \ref{wf_dyn2}. Adding the first-order correction allows states in the immediate next block (featuring only one ground state molecule with phonons) to contribute to the wavefunction. The timescale at which the first-order terms contribute is much longer than the ultrafast vibrational dynamics on each excited molecule. The exact wavefunction is recovered as one spans the entire Hilbert space from left to right in Fig. \ref{wf_dyn2}, but as can be appreciated, there are only a few molecules with phonons in the electronic ground state for large $N$, even for long times of interest, justifying the convenience of the shorthand notation in Eq. \ref{eq:renorm}.

Although the fact that there is only one molecular specie in the system is central for the permutational symmetries of Eqs. \ref{eq:perm1} and \ref{eq:perm2} to hold, addition of different molecular species can be done without dramatically increasing the computational cost since permutational symmetries still apply for each type. The renormalization of the coefficients associated to each species as in Eq. \ref{eq:renorm} will depend on their concentration. Similarly, disorder due to inhomogeneous broadening or spacial variation of the coupling to the photon mode, which has been shown to play a central role in molecular polaritonics systems  \cite{Schachenmayer2,Schachenmayer}, can be included by adding new molecular species for each value that is sampled according to the corresponding distribution of excitonic frequencies or interaction strength. Notice that the computational cost is still reduced significantly in this case since the distribution can be discretized into a small number of bins.
$\ $\\

\subsection{Zeroth-Order Approximation}\label{sec:z-o-a}

Let us assume that temperature $T=0$, meaning our initial state in the original notation is given by  $A^{(0)}_{111...1}(0)=1$. The time-dependent wavefunction (see Eq. \ref{eq:exact1}) at the zeroth-order approximation is given by the following vibrational wavefunctions
\begin{align}\label{eq:nophonon}
\psi^{(0)}(\vec{q},t)&=A^{(0)}_{11\cdots 1}(t)\prod_{k}^{N}\varphi_{1}(q_{k}) \ \ \ \ \textrm{and}\nonumber\\
\psi^{(i)}(\vec{q},t)&=\sum_{j_{i}}^{m}A^{(1)}_{j_{i}1\cdots 1}(t)\phi_{j_{i}}(q_{i})\prod_{k\neq i}^{N}\varphi_{1}(q_{k}).
\end{align} By using the renormalized permutationally-symmetric coefficients, the EoM in Eq. \ref{eq:EoM5} transform into those of an \textit{effective} single molecule strongly coupled to the cavity mode (with effective coupling $\sqrt{N}g$), 

\begin{widetext}

\begin{equation}\label{eq:Makri}
i\begin{pmatrix}
\dot{\tilde{A}}^{(0)}_{0}(t) \\
\dot{\tilde{A}}^{(1)}_{10}(t) \\
\dot{\tilde{A}}^{(1)}_{20}(t) \\
\vdots \\
\dot{\tilde{A}}^{(1)}_{m0}(t)
\end{pmatrix}=
\begin{pmatrix}
\omega_{c} & g\sqrt{N}\langle \varphi_{1}|\phi_{1}\rangle & g\sqrt{N}\langle \varphi_{1}|\phi_{2}\rangle & \cdots  & g\sqrt{N}\langle \varphi_{1}|\phi_{m}\rangle\\
g\sqrt{N}\langle \phi_{1}|\varphi_{1}\rangle & \omega_{eg,1} & 0 & \cdots  & 0\\
g\sqrt{N}\langle \phi_{2}|\varphi_{1}\rangle & 0 & \omega_{eg,2} & \cdots  & 0 \\
\vdots & \vdots  & \vdots & \ddots  & \vdots\\
g\sqrt{N}\langle \phi_{m}|\varphi_{1}\rangle & 0 & 0 & \cdots  & \omega_{eg,m}\\
\end{pmatrix} \begin{pmatrix}
\tilde{A}^{(0)}_{0}(t)\\
\tilde{A}^{(1)}_{10}(t)\\
\tilde{A}^{(1)}_{20}(t)\\
\vdots \\
\tilde{A}^{(1)}_{m0}(t)
\end{pmatrix}.
\end{equation}
\end{widetext}Eq. \ref{eq:Makri} is consistent with previous results where an impurity (in this case the optical mode) coupled to a large environment can be simplified into an impurity interacting with an effective harmonic bath that includes the relevant frequencies of the environment \cite{Makri}; in this case the optical transitions of an individual molecule. Alternatively, it is also consistent with the classical optics treatments of polaritons arising as the result of a photonic oscillator coupling to a set of effective oscillators representing the molecular transitions (e.g., transfer matrix methods) \cite{Schubert,Agranovich2}. This has also been justified by Keeling and co-workers within a quantum mechanical framework \cite{Cwik,KeelingZeb}. We can rewrite Eq. \ref{eq:Makri} in a form that is better suited for implementation in quantum dynamics packages such as the Multiconfiguration Time-Dependent Hartree (MCTDH) method \cite{mctdh,mctdh3,Vendrell2},
\begin{equation}\label{eq:Makri2}
i\begin{pmatrix}
\dot{\psi}_{ph}(q,t) \\
\dot{\psi}_{exc}(q,t) \\
\end{pmatrix}=
\begin{pmatrix}
\mathds{P}\hat{H}_{g}\mathds{P}+\omega_{c} & g\sqrt{N}\\
g\sqrt{N} & \hat{H}_{e}\\
\end{pmatrix}\begin{pmatrix}
\psi_{ph}(q,t) \\
\psi_{exc}(q,t) \\
\end{pmatrix}, 
\end{equation}with the projector over the ground vibrational state $\mathds{P}=|\varphi_{1}\rangle\langle \varphi_{1}|$ and the photonic and excitonic wavefunctions 

\begin{equation}
\psi_{ph}(q,t)=\tilde{A}^{(0)}_{0}(t)\varphi_{1}(q),\ \ \psi_{exc}(q,t)=\sum_{j}^{m}\tilde{A}_{j0}^{(1)}(t)\phi_{j}(q).
\end{equation} Here, we readily identify the zeroth-order Hamiltonian in Eq. \ref{eq:Makri2} as
\begin{align}\label{eq:heff1}
\hat{\tilde{H}}^{(0)}&=\left(\mathds{P}\hat{H}_{g}\mathds{P}+\omega_{c}\right)|1\rangle\langle 1|+\hat{H}_{e}|e\rangle\langle e|\nonumber\\
&+g\sqrt{N}\left(|e\rangle\langle 1|+|1\rangle\langle e|\right),
\end{align}with $\hat{H}_{g/e}=\frac{1}{2\mu}\hat{p}^{2}+V_{g/e}(\hat{q})$. Eq. \ref{eq:Makri2} can be used to recover the original time-dependent many-body wavefunctions in Eq. \ref{eq:nophonon},

\begin{align}
&\psi^{(0)}(\vec{q},t)=\psi_{ph}(q_{1},t)\prod_{k=2}^{N}\varphi_{1}(q_{k}),\nonumber\\
&\psi^{(i)}(\vec{q},t)=\frac{1}{\sqrt{N}}\psi_{exc}(q_{i},t)\prod_{k\neq i}^{N}\varphi_{1}(q_{k}).
\end{align}

As far as we are aware, the separation between collective and single-molecule emission processes was first pointed out by Spano when calculating vibrationally-dressed lower polariton states \cite{Spano2}, although crucial ideas were introduced by the same author much earlier \cite{Spano3}, as we shall discuss later. In a more recently article, the same author combined such ideas with the permutationally symmetric basis to compute photoluminescent spectra starting from approximated vibro-polaritonic eigenstates \cite{Spano}. Our work formalizes Spano's observations into the hierarchy summarized in Fig. \ref{wf_dyn2}, and capitalizes it to compute dynamics in complex molecular polaritonic systems featuring arbitrary PESs.

Two important comments are in order concerning Eq. \ref{eq:heff1}. First, the artificial effective molecule in the ground state is only allowed to be in its ground vibrational state (no phonons). The physical intuition for this is simple: imagine the molecules have been collectively excited at their Franck-Condon (FC) configurations by the cavity field; each of them can in principle re-emit such energy into the cavity creating any vibrational state of their ground electronic state. However, only when they go back to their \textit{vibrational} ground state, the number of ground state molecules with phonons is conserved and the process becomes collective. This seemingly unremarkable observation has an important implication: it states that single-molecule polariton simulations are not applicable to the collective regime unless the light-matter coupling is restricted to the FC region, where the vibrational ground state $\varphi_{1}(q)$ has significant amplitude (see Fig. \ref{FC}). However, as Eq. \ref{eq:Makri2} reveals, such simulations can be simply adjusted for the collective regime by the appropriate inclusion of the projector $\mathds{P}$. This projector can be readily implemented in various ways (e.g., by restricting the vibrational basis in the ground state to $|\varphi_{1}\rangle$ or by projecting the light-matter coupling, among several options). Second, the results of this section are exact in the thermodynamic ($N\rightarrow\infty$) limit because they imply $g\rightarrow 0$. We next add the first-order correction, where the wavefunction now has states with amplitude of the order of $g$.

\subsection{First-Order Correction and Beyond}

Proceeding analogously, the EoM \textit{up to} first order can be written as

\begin{widetext}
\vspace{-0.2in}
\begin{equation*}
i\begin{pmatrix}
\dot{\tilde{A}}^{(0)}_{0}(t) \\
\dot{\tilde{A}}^{(1)}_{10}(t) \\
\dot{\tilde{A}}^{(0)}_{10}(t) \\
\dot{\tilde{A}}^{(1)}_{11}(t) \\
\end{pmatrix}=\begin{pmatrix}
\omega_{c} & g\sqrt{N}\langle \varphi_{1}|\phi_{1}\rangle  & 0 & 0 \\
g\sqrt{N}\langle \phi_{1}|\varphi_{1}\rangle & \omega_{eg,1} &  \boxed{g\langle \phi_{1}|\varphi_{2}\rangle} & 0\\
0 & \boxed{g\langle \varphi_{2}|\phi_{1}\rangle} & \omega_{g,2}+\omega_{c} & g\sqrt{N-1}\langle \varphi_{1}|\phi_{1}\rangle \\
0 & 0 & g\sqrt{N-1}\langle \phi_{1}|\varphi_{1}\rangle & \omega_{g,2}+\omega_{eg,1}\\
\end{pmatrix}\begin{pmatrix}
\tilde{A}^{(0)}_{0}(t) \\
\tilde{A}^{(1)}_{10}(t) \\
\tilde{A}^{(0)}_{10}(t) \\
\tilde{A}^{(1)}_{11}(t) \\
\end{pmatrix},
\end{equation*}
\end{widetext}where we considered the case $m=1$ for illustration purposes.  The general case can be written in a compact form using projection operators
\begin{widetext}
\vspace{-0.2in}
\begin{align}\label{eq:1oapp}
\hat{\tilde{H}}^{(1)}&=\left(\mathds{P}_{1}\hat{H}_{g,1}\mathds{P}_{1}+\mathds{P}_{2}\hat{H}_{g,2}\mathds{P}_{2}+\omega_{c}\right)|1,0\rangle\langle 1,0|+\left(\hat{H}_{e,1}+\mathds{P}_{2}\hat{H}_{g,2}\mathds{P}_{2}\right)|e,0\rangle\langle e,0|\nonumber\\
&+\left(\mathds{Q}_{1}\hat{H}_{g,1}\mathds{Q}_{1}+\mathds{P}_{2}\hat{H}_{g,2}\mathds{P}_{2}+\omega_{c}\right)|1,1\rangle\langle 1,1|+\left(\mathds{Q}_{1}\hat{H}_{g,1}\mathds{Q}_{1}+\hat{H}_{e,2}\right)|e,1\rangle\langle e,1|\nonumber\\
&+g\sqrt{N}\left(|e,0\rangle\langle 1,0|+|1,0\rangle\langle e,0|\right)+g\sqrt{N-1}\left(|e,1\rangle\langle 1,1|+|1,1\rangle\langle e,1|\right)+g\left(|e,0\rangle\langle 1,1|+|1,1\rangle\langle e,0|\right),
\end{align}
\end{widetext}where $\mathds{Q}_{i}=\mathds{1}_{vib,i}-\mathds{P}_{i}$. When compared to $\hat{H}^{(0)}$ in Eq. \ref{eq:heff1}, we notice the emergence of a second label that corresponds to the number of ground state molecules with phonons. Terms involving the $|1,0\rangle$ and $|e,0\rangle$ states correspond to the zeroth-order approximation and describe the aforementioned effective molecule (say, molecule $1$) coupled to the cavity with amplitude $\sqrt{N}g$. Similar terms involve coupling the $|1,1\rangle$ and $|e,1\rangle$ states (corresponding to molecule $2$) with a slightly weaker amplitude $\sqrt{N-1}g$, but \textit{only} when molecule $1$ has phonons in the ground state. This effective light-matter interaction nonlinearity, which conditions the light-matter interaction of the second molecule on the presence of phonons in the first one, might look odd at first sight, but must be endorsed as the emergent physics resulting from the hierarchy. As in the previous section, $\hat{\tilde{H}}^{(1)}$ can be written in a more suitable form for implementation in wavepacket dynamics methods
\begin{widetext}
{\scriptsize
\begin{equation}\label{eq:wpd1oc}
i\begin{pmatrix}
\dot{\psi}_{ph}^{(0)}(q_{1},q_{2})\\
\dot{\psi}_{exc}^{(0)}(q_{1},q_{2})\\
\dot{\psi}_{ph}^{(1)}(q_{1},q_{2})\\
\dot{\psi}_{exc}^{(1)}(q_{1},q_{2})\\
\end{pmatrix}=\begin{pmatrix}
\mathds{P}_{1}\hat{H}_{g,1}\mathds{P}_{1}+\mathds{P}_{2}\hat{H}_{g,2}\mathds{P}_{2}+\omega_{c}& g\sqrt{N} & 0 & 0 \\
g\sqrt{N} & \hat{H}_{e,1}+\mathds{P}_{2}\hat{H}_{g,2}\mathds{P}_{2} & \boxed{g} & 0\\
0 & \boxed{g} & \mathds{Q}_{1}\hat{H}_{g,1}\mathds{Q}_{1}+\mathds{P}_{2}\hat{H}_{g,2}\mathds{P}_{2}+\omega_{c} & g\sqrt{N-1}\\
0 & 0 & g\sqrt{N-1} &\mathds{Q}_{1}\hat{H}_{g,1}\mathds{Q}_{1}+\hat{H}_{e,2}\\
\end{pmatrix} \begin{pmatrix}
\psi_{ph}^{(0)}(q_{1},q_{2})\\
\psi_{exc}^{(0)}(q_{1},q_{2})\\
\psi_{ph}^{(1)}(q_{1},q_{2})\\
\psi_{exc}^{(1)}(q_{1},q_{2})\\
\end{pmatrix}.
\end{equation}}
\end{widetext}Notice that ignoring the matrix elements proportional to $g$ (boxed) amounts to block-diagonalizing the Hamiltonian according to the number of electronic ground state molecules with phonons, reiterating the approximate symmetry featured in Fig. \ref{wf_dyn2}. Eq. \ref{eq:wpd1oc} can be interpreted as two effective molecules coupled to the cavity. Finally, the original many-body wavefunction in Eqs. \ref{eq:exact2} and \ref{eq:exact3} can be rewritten in terms of the 2-molecule wavepackets as

\begin{align}
\psi^{(0)}(\vec{q},t)&=\psi^{(0)}_{ph}(q_{1},q_{2},t)\prod_{k=3}^{N}\varphi_{1}(q_{k})\nonumber\\
&+\sum_{i}^{N}\frac{\psi^{(1)}_{ph}(q_{i},q_{i'\neq i},t)}{\sqrt{N}}\prod_{k\neq i,i'}^{N}\varphi_{1}(q_{k}),\nonumber
\end{align}
\begin{align}
\psi^{(i)}(\vec{q},t)&=\frac{\psi^{(0)}_{exc}(q_{i},q_{i'\neq i},t)}{\sqrt{N}}\prod_{k\neq i,i'}^{N}\varphi_{1}(q_{k})\nonumber\\
&+\sum_{j\neq i}^{N}\frac{\psi^{(1)}_{exc}(q_{j\neq i},q_{i},t)}{\sqrt{N(N-1)}}\prod_{k\neq i,j}^{N}\varphi_{1}(q_{k}),
\end{align} with $i'$ is an arbitrary molecule of the ensemble and 

\begin{align}
&\psi^{(0)}_{ph}(q_{1},q_{2},t)=\tilde{A}^{(0)}_{0}(t)\varphi_{1}(q_{1})\varphi_{1}(q_{2}),\nonumber\\
&\psi^{(1)}_{ph}(q_{1},q_{2},t)=\sum_{k>1}^{m}\tilde{A}^{(0)}_{k}(t)\varphi_{k}(q_{1})\varphi_{1}(q_{2}),\nonumber\\
&\psi^{(0)}_{exc}(q_{1},q_{2},t)=\sum_{l}^{m}\tilde{A}^{(1)}_{l0}(t)\phi_{l}(q_{1})\varphi_{1}(q_{2}),\nonumber\\
&\psi^{(1)}_{exc}(q_{1},q_{2},t)=\sum_{l,k>1}^{m}\tilde{A}^{(1)}_{lk}(t)\phi_{l}(q_{2})\varphi_{k}(q_{1}).
\end{align} We make use of this Hamiltonian in Section \ref{sec:relax} and show that adding the first-order correction describes collective processes, in addition to those whose rates scale as $1/N$. Similarly, corrections of $k$th order require using a Hamiltonian of $k+1$ effective molecules interacting with a cavity mode, and would include information about phenomena with rates that scale up to $1/N^{k}$. Such decomposition of the many-molecules problem into a few-molecules problem is what we call Collective dynamics Using Truncated Equations (CUT-E) method.

Let us summarize the central results of the article. Application of the permutational symmetries in Eqs. \ref{eq:perm1} and \ref{eq:perm2} to the Time-Dependent Schr\"odinger Equation for $N$ identical molecules in a cavity gives rise to the EoM in Eq. \ref{eq:EoM5}. These EoM are endowed with a convenient hierarchy of timescale separations, schematically illustrated by the scissors in Fig. \ref{wf_dyn2}. For a fixed collective coupling $\sqrt{N}g$, processes with $N$-independent ($\mathcal{O}(N^{-1})$) rates are captured by a single (two) effective molecule(s) in a cavity, according to Eq. \ref{eq:Makri2} (Eq. \ref{eq:wpd1oc}). This is schematically represented in Fig. \ref{micro}b. Although high order corrections might become relevant when a small number of molecules is considered, here we are interested in the case where a large number of molecules is used to reach collective strong light-matter coupling. Thus, we will defer the exploration of corrections beyond first order for future works.

\section{Observables in the zeroth-order approximation}\label{sec:zero}

Observables of the real system must be calculated using the many-body wavefunction $|\Psi(t)\rangle$ and not directly using the effective single-molecule wavefunction $|\tilde{\Psi}(t)\rangle$. However, there are particular observables for which both wavefunctions provide the same answer. This will depend on whether the observable is local or collective. For pedagogical purposes, throughout this section, we assume the zeroth-order approximation.

\subsection{Local Properties}

Hereafter, we define local properties to be quantities that only depend on one single molecule (e.g. a chemical reaction) or the photon field alone.

\subsubsection{Chemical Properties:}

Consider a local observable $\hat{\Omega}^{(i)}$ that depends only on degrees of freedom of molecule $i$. As an example, assume we are interested in the nuclear dynamics in the excited state $\hat{\Omega}^{(i)}=\hat{q}_{i}|e_{i}\rangle\langle e_{i} |$. Using the zeroth-order wavefunction in Eq. \ref{eq:nophonon} we obtain

\begin{equation}
\langle \Psi(t)|\hat{q}_{i}|e_{i}\rangle\langle e_{i}|\Psi(t)\rangle=\frac{1}{N}\langle \tilde{\Psi}(t)|\hat{q}|e\rangle\langle e|\tilde{\Psi}(t)\rangle.
\end{equation} This is a remarkable result because it demonstrates that the excited-state dynamics of $N$ molecules collectively coupled to a cavity mode (described by $|\Psi(t)\rangle$) is identical to the dynamics of an effective single molecule strongly coupled to a cavity (described by $|\tilde{\Psi}(t)\rangle$), except for a constant $1/N$ dilution factor. This factor is just a consequence of using a single photon to alter the excited-state dynamics of $N$ molecules. The probability of \textit{any} molecule being at the configuration $q$ is the same as that of the single effective molecule strongly coupled to the cavity. However, we emphasize that the effective molecule is not allowed to have phonons while it is in the ground state, a restriction which can introduce significant differences compared to standard single-molecule calculations.

An important corollary of the above analysis is that effects predicted using single-molecule models might actually occur in the collective regime if they rely on changes in the excited PES at the FC region, but not the ones relying on changes beyond. This was first pointed out by Galego and coworkers \cite{Feist1} by analyzing polaritonic PESs for an ensemble of molecules interacting with a cavity mode. This is also consistent with a recent work by Cui and Nitzan \cite{Nitzan}, who concluded that excited-state dynamics in polaritonic systems is dominated by states that are reachable from the ground electronic state. Yet, we propose that modifications of chemical dynamics that occur on a timescale longer than the decay of the initially prepared excitations can occur in an $N$-independent manner if they dramatically depend on the ultrafast polariton-modified dynamics at the FC region. In section \ref{sec:nonstat} we present such an example.

\subsubsection{Optical properties:}

Another set of properties that are equivalent in the single molecule and $N\rightarrow \infty$ cases are those that can be extracted only from the dynamics of the field. For example, the linear transmission, absorption and reflection spectra can be calculated from the photon autocorrelation function $c(t)$. In the zeroth-order approximation, it can be shown that $c(t)=\langle \Psi(0) | \Psi(t) \rangle=\langle \tilde{\Psi}(0) |\tilde{\Psi}(t) \rangle$, where $|\tilde{\Psi}(0)\rangle=\varphi_{0}(q)|1\rangle$ is the photonic state. This is consistent with previous work by the Keeling group \cite{Cwik,KeelingZeb}.

\subsection{Non-Local Properties}

Another set of observables consists of operators that are delocalized across several molecules or depend on both molecular and optical degrees of freedom. Some of the most common observables of this kind are the populations of the polariton and dark-states. Such observables can be obtained with the effective single-molecule model but not from its reduced electronic-photonic density matrix, as we will show next. 

\subsubsection{Polariton and Dark-States Populations}\label{sec:padsp}

\begin{figure}[htp]
\begin{center}
\includegraphics[width=0.8\linewidth]{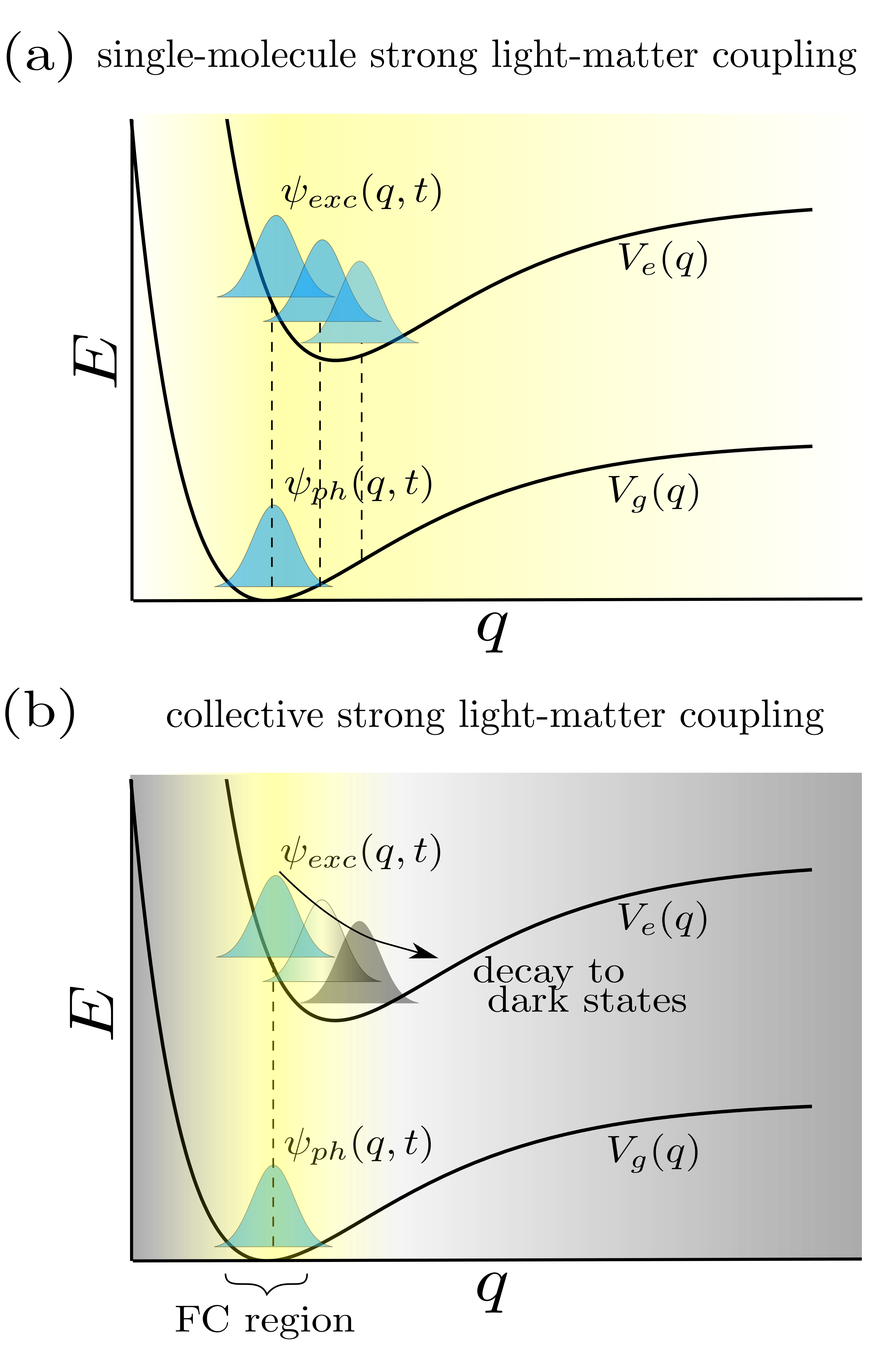}
\caption{(a) For a cavity containing a single molecule, light-matter coupling (denoted in yellow) has the ability to modify nuclear dynamics throughout all configurations $q$. (b) This situation contrasts with a cavity containing an infinite number of molecules (zeroth-order approximation), where collective light-matter coupling is localized at the Franck Condon (FC) region. Nuclear displacement away from this region is equivalent to decay of polaritons into dark states.}
\label{FC}
\end{center}
\end{figure}

Let us write the expectation value of an arbitrary electronic-photonic operator $\langle\hat{\Omega}\rangle=\textrm{Tr}\left[\hat{\rho}\hat{\Omega}\right] $ in terms of the reduced density matrix of the effective single-molecule system $\hat{\tilde{\rho}}$:

\begin{align}\label{eq:obs}
\langle \hat{\Omega}\rangle&=\textrm{Tr}\left[\hat{\rho}\hat{\Omega} \right]=\sum_{ij}\rho_{ij}\Omega_{ji}=\tilde{\rho}_{11}\Omega_{11}+\tilde{\rho}_{1e}\sum_{i=1}^{N}\frac{1}{\sqrt{N}}\Omega_{1e_{i}}\nonumber\\
&+\tilde{\rho}_{e1}\sum_{i=1}^{N}\frac{1}{\sqrt{N}}\Omega_{e_{i}1}+\tilde{\rho}_{ee}\frac{1}{N}\sum_{i=1}^{N}\Omega_{e_{i}e_{i}}\nonumber\\
&+\frac{1}{N}\langle \tilde{\Psi}(t)|\tilde{\Psi}_{FC} \rangle\langle\tilde{\Psi}_{FC} |\tilde{\Psi}(t) \rangle \left(\sum_{i,i'\neq i}\Omega_{e_{i}e_{i'}} \right),
\end{align}where we have identified $|\tilde{\Psi}_{FC}\rangle=\varphi_{1}(q)|e\rangle$ as the FC wavepacket. The above equation implies that the reduced density matrix $\tilde{\rho}$ of the effective single molecule is, in general, not enough to calculate any delocalized molecular observables since the last term of Eq. \ref{eq:polpop} refers to the projection of the wavefunction onto the FC wavepacket, which corresponds to the inter-exciton coherences: 

\begin{equation}\label{eq:expec}
\langle e_{i} |\hat{\rho}(t)|e_{i'}\rangle=\frac{1}{N}\langle  \tilde{\Psi}(t)| \tilde{\Psi}_{FC} \rangle \langle \tilde{\Psi}_{FC}| \tilde{\Psi}(t)\rangle.
\end{equation} An example, let $\hat{\Omega}=|P\rangle\langle P|$, with $|P\rangle=c_{0}|1\rangle+c_{1}|B\rangle$ (where $|B\rangle=\frac{1}{\sqrt{N}}\sum_{i=1}^{N}|e_{i}\rangle$ is the totally-symmetric excitonic state) being one of the polariton states. Using Eq. \ref{eq:obs}, population of this state is calculated to be 
\begin{align}\label{eq:polpop}
&\langle \Psi(t)| P\rangle\langle P| \Psi(t) \rangle \approx|c_{0}|^{2}\tilde{\rho}_{11}(t)+c^{*}_{0}c_{1}\tilde{\rho}_{1e}+c^{*}_{1}c_{0}\tilde{\rho}_{e1}\nonumber\\
&+|c_{1}|^{2}\langle \tilde{\Psi}(t)|\tilde{\Psi}_{FC} \rangle\langle\tilde{\Psi}_{FC} |\tilde{\Psi}(t) \rangle.
\end{align}  Similarly, dark states can be written as $|D\rangle=\sum_{i}c_{i}|e_{i}\rangle$, with $|c_{1}|=|c_{2}|=\cdots=|c_{N}|=1/\sqrt{N}$, $\sum_{i=1}^{N}c_{i}=0$, and $\Omega_{e_{i}e_{j}}=\langle e_{i}|D\rangle\langle D |e_{j}\rangle = c_{i}c^{*}_{j}$ (i.e., they are chosen orthogonal to the $|P\rangle$ states, so here we take them to be the Fourier combinations of excitons that are orthogonal to $|B\rangle$; see for instance \cite{Raphael}). This leads to

\begin{equation}\label{eq:obs2}
\langle \Psi(t) |D\rangle\langle D | \Psi(t) \rangle=\frac{1}{N}\langle \tilde{\Psi}(t)|e\rangle\left( \mathds{1}-|\varphi_{1}\rangle\langle\varphi_{1}|\right)\langle e|\tilde{\Psi}(t)\rangle.
\end{equation} Notice that this calculation is identical for every dark state in the chosen basis, therefore the population in the dark-state manifold yields

\begin{align}\label{eq:dspop}
\sum_{k}^{N-1}\langle |D\rangle\langle D | \rangle&\approx 1-\langle \tilde{\Psi}(t)| 1 \rangle\langle 1 |\tilde{\Psi}(t) \rangle\nonumber\\
&-\langle \tilde{\Psi}(t)|\tilde{\Psi}_{FC} \rangle\langle\tilde{\Psi}_{FC} |\tilde{\Psi}(t) \rangle,
\end{align}where we have made $\frac{N-1}{N}\approx 1$ since the zeroth-order approximation becomes exact for $N\rightarrow \infty$. From this result, we can extract the intuitive interpretation that the bright state $|B\rangle$ in the $N$-molecule system corresponds to a FC wavepacket in the excited state of the effective single molecule, while the dark states correspond to the rest of the wavefunction whose nuclear configurations lie outside of the FC region (See Fig. \ref{FC}). 

For cases where the chemically relevant excited-state dynamics involves fast changes of the nuclear configurations away from the FC point, excited state reactivity is essentially relaxation to dark-states. The question is whether the relaxation occurs along the reactive coordinate of interest (e.g., particular reactive dark-modes), or whether it occurs along orthogonal modes to it. Thus, we have derived a powerful design principle for polariton chemistry which has so far unjustifiably gathered little attention: the strategy is not to avoid decay into dark states, which seems inexorable in most cases, but to use strong coupling to control which dark states to target. In fact, using the zeroth-order approximation, we will illustrate some mechanisms to manipulate ratios for these relaxation pathways in Section \ref{sec:nonstat}.

\section{Polariton Vibrational Relaxation}\label{sec:relax}

Previous work by del Pino and coworkers \cite{delPino} addressed the relaxation dynamics of molecular polaritons using the Hamiltonian in Eq. \ref{eq:ham} for a vibrational harmonic bath and linear vibronic coupling,

\begin{align}
\hat{H}_{m}^{(i)}&=\sum_{k}\omega_{\nu,k}\hat{b}_{k}^{(i)\dagger}\hat{b}^{(i)}_{k}\nonumber\\
&+\left[\omega_{eg,1}+\sum_{k}\omega_{\nu,k}\sqrt{s_{k}}(\hat{b}_{k}^{(i)\dagger}+\hat{b}^{(i)}_{k})\right]|e_{i}\rangle\langle e_{i}|.
\end{align} This model of relaxation is the single-photon mode simplification of a previous model by Litinskaya and Agranovich \cite{agranovich}. In the limit where vibronic coupling is much smaller than the collective light-matter coupling and there is an energy-dense set of vibrational modes $k$, we can use our formalism to analytically derive relaxation rates by including a collection of local vibrational modes and using Fermi's Golden Rule with the vibronic couplings as the perturbation. Although such a result is already well known, it serves as a benchmark and illustration for our formalism. To unclutter the calculations, let us consider the case where the excitons and the cavity are in resonance ($\omega_{c}=\omega_{eg,1}=\omega$).

\subsection{Zeroth-Order Approximation: $N$-Independent Effects}

\begin{figure}[htb]
\begin{center}
\includegraphics[width=1\linewidth]{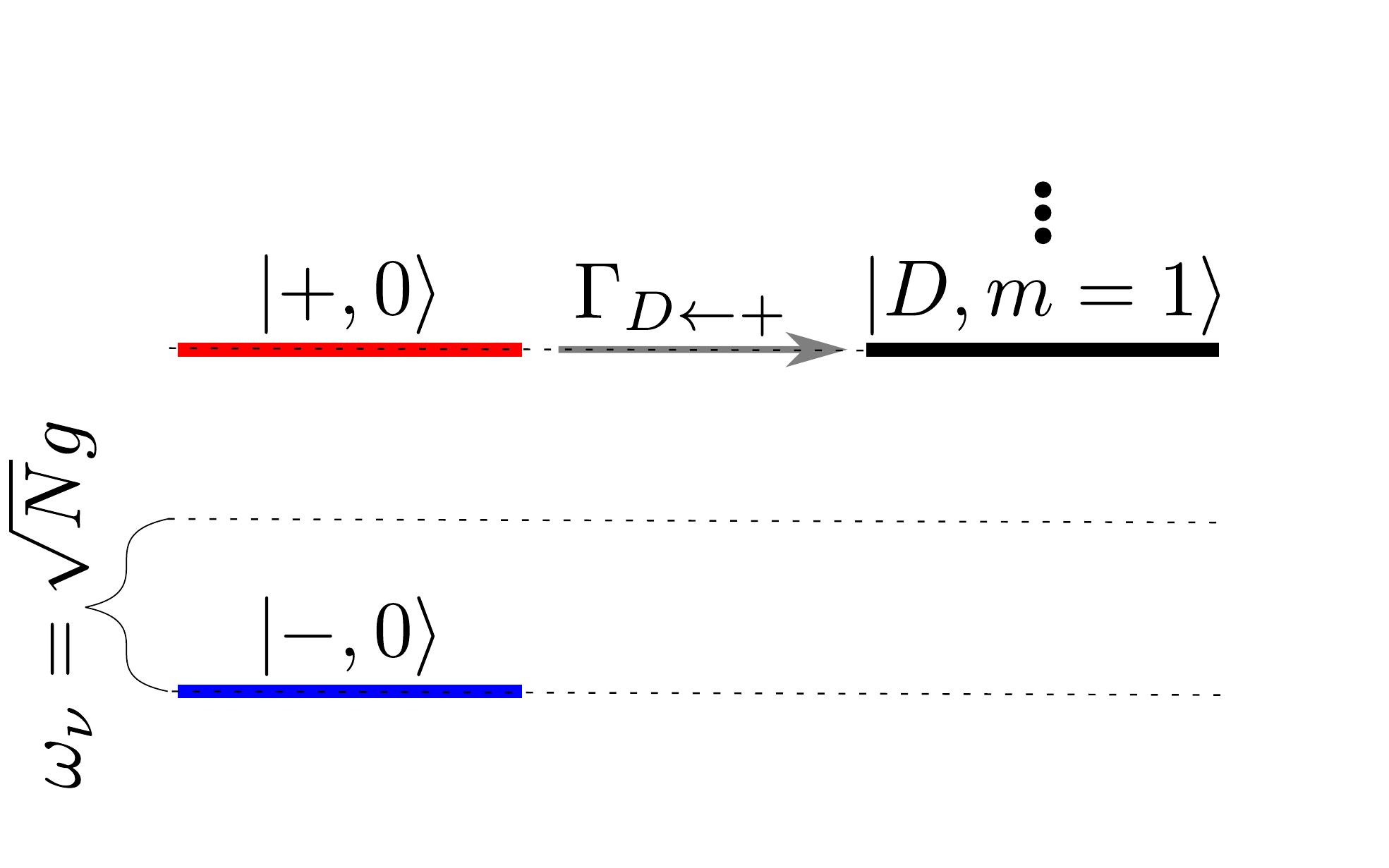}
\caption{Relaxation from the upper polariton to the dark states as described by the zeroth-order approximation (blue: lower polariton, red: upper polariton, black: dark states). States $|+,0\rangle$ and $|D,m=1\rangle$ are resonantly coupled through vibronic coupling. $m=1$ means there is 1 phonon in one of the bath modes of the excited effective molecule.}
\label{decay0}
\end{center}
\end{figure}

Using Eq. \ref{eq:heff1}, the model of relaxation in the zeroth-order approximation is given by the unperturbed and vibronic coupling Hamiltonians
\begin{equation*}
\hat{\tilde{H}}^{(0)}=\hat{\tilde{H}}^{(0)}_{0} +\hat{\tilde{H}}^{(0)}_{I},
\end{equation*}
\begin{align*}
\hat{\tilde{H}}^{(0)}_{0}&=\omega|1,0\rangle\langle 1,0|+\left(\omega+\sum_{k}\omega_{\nu,k}\hat{b}_{k}^{\dagger}\hat{b}_{k}\right)|e\rangle\langle e|\nonumber\\
&+\sqrt{N}g\left(|e,0\rangle\langle 1,0 |+|1,0\rangle\langle e,0 | \right),
\end{align*}
\begin{equation}
\hat{\tilde{H}}^{(0)}_{1}=\sum_{k}\omega_{\nu,k}\sqrt{s_{k}}\left(\hat{b}_{k}^{\dagger}+\hat{b}_{k}\right)|e\rangle \langle e|,
\end{equation}where we have used the Fock basis for the vibrational bath (the second index ``$0$" means all vibrational modes $k$ are empty; do not confuse this notation with that of Eq. \ref{eq:1oapp}). The eigenstates of $\hat{\tilde{H}}^{(0)}_{0}$ are trivial,
\begin{align}
& |\pm,0\rangle=\frac{1}{\sqrt{2}}\left(|e,0\rangle\pm |1,0\rangle \right)\nonumber\\
& |D,m\rangle=|e,m> 0\rangle,
\end{align}where $m>0$ denotes that at least one mode of the vibrational bath is not in the vacuum state. The eigenvalues are given by $\omega_{\pm,0}=\omega\pm g$ and $\omega_{D,m}=\omega+\sum_{k}\omega_{\nu,k}m_{k}$ respectively. Using Fermi's Golden Rule we can obtain the following relaxation rates:
\begin{align}\label{eq:uptods1}
\Gamma_{D \leftarrow +}&=2\pi\sum_{m}|\langle D,m |\hat{\tilde{H}}^{(0)}_{1}|+,0\rangle|^{2}\delta(\omega_{D,m}-\omega_{+,0})\nonumber\\
&=\frac{2\pi}{2}\sum_{k}\omega_{\nu,k}^{2}s_{k}\delta(g-\omega_{\nu,k}),
\end{align}
\begin{equation}\label{eq:uptolp1}
\Gamma_{- \leftarrow +}=2\pi|\langle -,0|\hat{\tilde{H}}^{(1)}|+,0\rangle|^{2} \delta(2g)=0,
\end{equation}
\begin{equation}\label{eq:dtolp1}
\Gamma_{- \leftarrow D}=2\pi|\langle -,0|\hat{\tilde{H}}^{(1)}|e,m\rangle|^{2} \delta(2g-\sum_{k}\omega_{\nu,k}m_{k})=0,
\end{equation} where the last two rates are equal to $0$ because the resonance condition is not fulfilled. A schematic representation of the relaxation dynamics is shown in Fig. \ref{decay0}.

Apart from missing a factor of $\frac{N-1}{N}$ in $\Gamma_{D \leftarrow +}$, the zeroth-order approximation correctly predicts the relaxation rate from upper polariton to dark states (which in this weak-vibronic coupling model in the effective single-molecule calculation, corresponds to the bare molecular exciton with a single phonon $|e,1\rangle$), but disregards both the upper to lower polariton, and dark state to lower polariton rates \cite{delPino}. This makes sense since the latter two are known to be proportional to $1/N$, and the zeroth-order approximation is exact for $N\rightarrow \infty$. Thus, for this simple model, the only relaxation process of relevance at ultrafast timescales is the downhill one \textit{into} dark states. 

\subsection{First-Order Correction: $1/N$ Effects}

By analogy with the previous section, we describe the vibrations using the eigenbasis of the electronic ground-state vibrational Hamiltonians for the two molecules. Adding the first-order correction Eq. \ref{eq:1oapp}, the relaxation model is described by the Hamiltonians

\begin{widetext}

\begin{align}
&\hat{\tilde{H}}^{(1)}=\hat{\tilde{H}}^{(1)}_{0} +\hat{\tilde{H}}^{(1)}_{1},\nonumber\\
&\hat{\tilde{H}}^{(1)}_{0}=\omega|1,0,0,0\rangle\langle 1,0,0,0|+\sum_{m}\left(\omega+\sum_{k}\omega_{\nu,k}m_{k}\right)|e,0,m,0\rangle\langle e,0,m,0|+\sum_{m>0}\left(\omega+\sum_{k}\omega_{\nu,k}m_{k}\right)|1,1,m,0\rangle\langle 1,1,m,0|\nonumber\\
&+\sum_{m>0,n}\left(\omega+\sum_{k}\omega_{\nu,k}m_{k}+\sum_{k}\omega_{\nu,k}n_{k}\right)|e,1,m,n\rangle\langle e,1,m,n|+g\sqrt{N}\left(|1,0,0,0\rangle\langle e,0,0,0|+|e,0,0,0\rangle\langle 1,0,0,0|\right)\nonumber\\
&+g\sqrt{N-1}\sum_{m>0}\left(|1,1,m,0\rangle\langle e,1,m,0|+|e,1,m,0\rangle\langle 1,1,m,0|\right)+g\sum_{m>0}\left(|e,0,m,0\rangle\langle 1,1,m,0|+|1,1,m,0\rangle\langle e,0,m,0|\right),\nonumber\\
&\hat{\tilde{H}}^{(1)}_{1}=\sum_{k}\omega_{\nu,k}\sqrt{s_{k}}\left(\hat{b}^{\dagger}_{1,k} + \hat{b}_{1,k}\right)|e,0\rangle\langle e,0|+\sum_{k}\omega_{\nu,k}\sqrt{s_{k}}\left(\hat{b}^{\dagger}_{2,k} + \hat{b}_{2,k}\right)|e,1\rangle\langle e,1|,
\end{align}
\end{widetext}where the last two labels in the kets represent the vibrational states of molecules $1$ and $2$ in the aforementioned Fock basis, $\hat{b}_{i,k}$ is the annihilation operator for the vibrational excitations of molecule $i$, and $m>0$ means $m_{k}>0$ for at least one mode $k$. The eigenstates of $\hat{\tilde{H}}^{(1)}_{0}$ are given by
{\small
\begin{align*}
&|\pm,0,0\rangle=\frac{1}{\sqrt{2}}\left(|1,0,0,0\rangle\pm |e,0,0,0\rangle\right),\\
&|\pm,m>0,0\rangle=\frac{1}{\sqrt{2}}|1,1,m>0,0\rangle\nonumber\\
&\pm \frac{1}{\sqrt{2}}\left(\sqrt{\frac{N-1}{N}}|e,1,m>0,0\rangle+\frac{1}{\sqrt{N}}|e,0,m>0,0\rangle\right),\\
&|D,m>0,0\rangle=\frac{1}{\sqrt{N}}|e,1,m>0,0\rangle+\sqrt{\frac{N-1}{N}}|e,0,m>0,0\rangle,\\
&|D,m>0,n\rangle=|e,1,m>0,n>0\rangle,
\end{align*}}with eigenvalues $\omega_{\pm,0,0}=\omega\pm g\sqrt{N}$, $\omega_{\pm,m,0}=\omega+\sum_{k}\omega_{\nu,k}n_{k}\pm g\sqrt{N}$, and $\omega_{D,m,n}=\omega+\sum_{k}\omega_{\nu,k}m_{k}+\sum_{k}\omega_{\nu,k}n_{k}$ respectively.  We can use these states to calculate the rate from the upper polariton to the lower polariton using Fermi's Golden Rule to get

\begin{figure*}[htb]
\begin{center}
\includegraphics[width=1\linewidth]{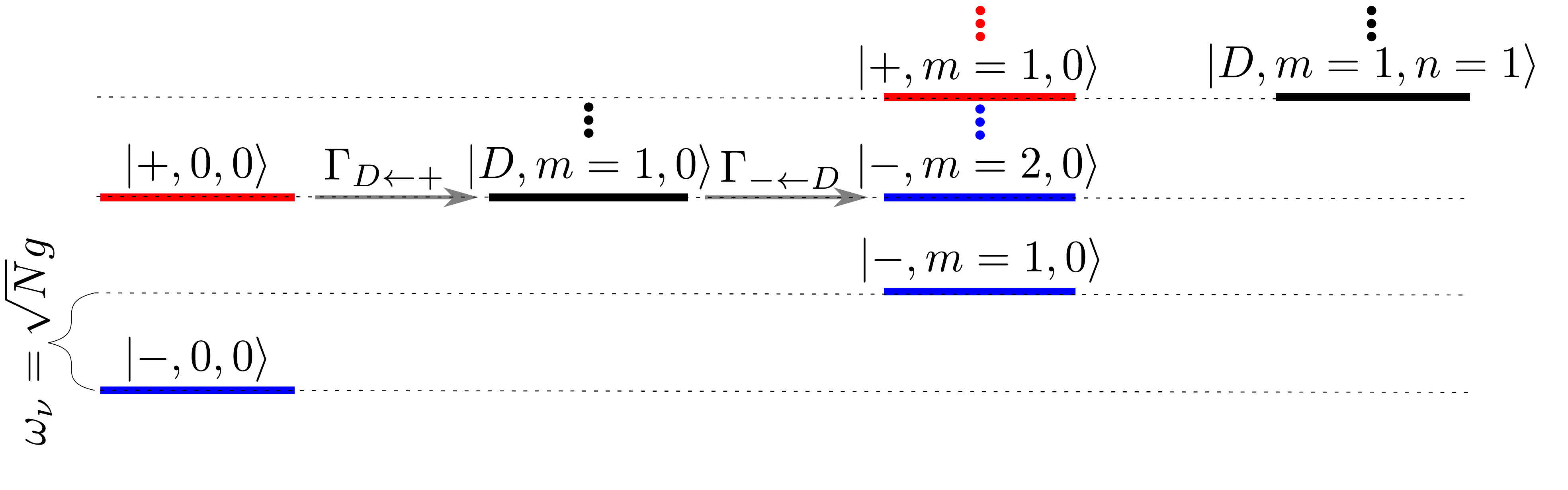}
\caption{Relaxation from the upper polariton to the lower polariton passing through the dark states, as described by inclusion of the first-order correction (blue: lower polariton, red: upper polariton, black: dark states). States $|+,0,0\rangle$, $|D,m=1,0\rangle$, and $|-,m=2,0\rangle$ are resonantly coupled through vibronic coupling. $m=2$ means there are 2 phonons of frequency $\omega_{\nu}=\sqrt{N}g$ in the bath modes of the effective molecule $1$. In the derivation, we have assumed the second phonon is emitted into a different bath mode than the first one.}
\label{decay1}
\end{center}
\end{figure*}

\begin{figure*}[htb]
\begin{center}
\includegraphics[width=1\linewidth]{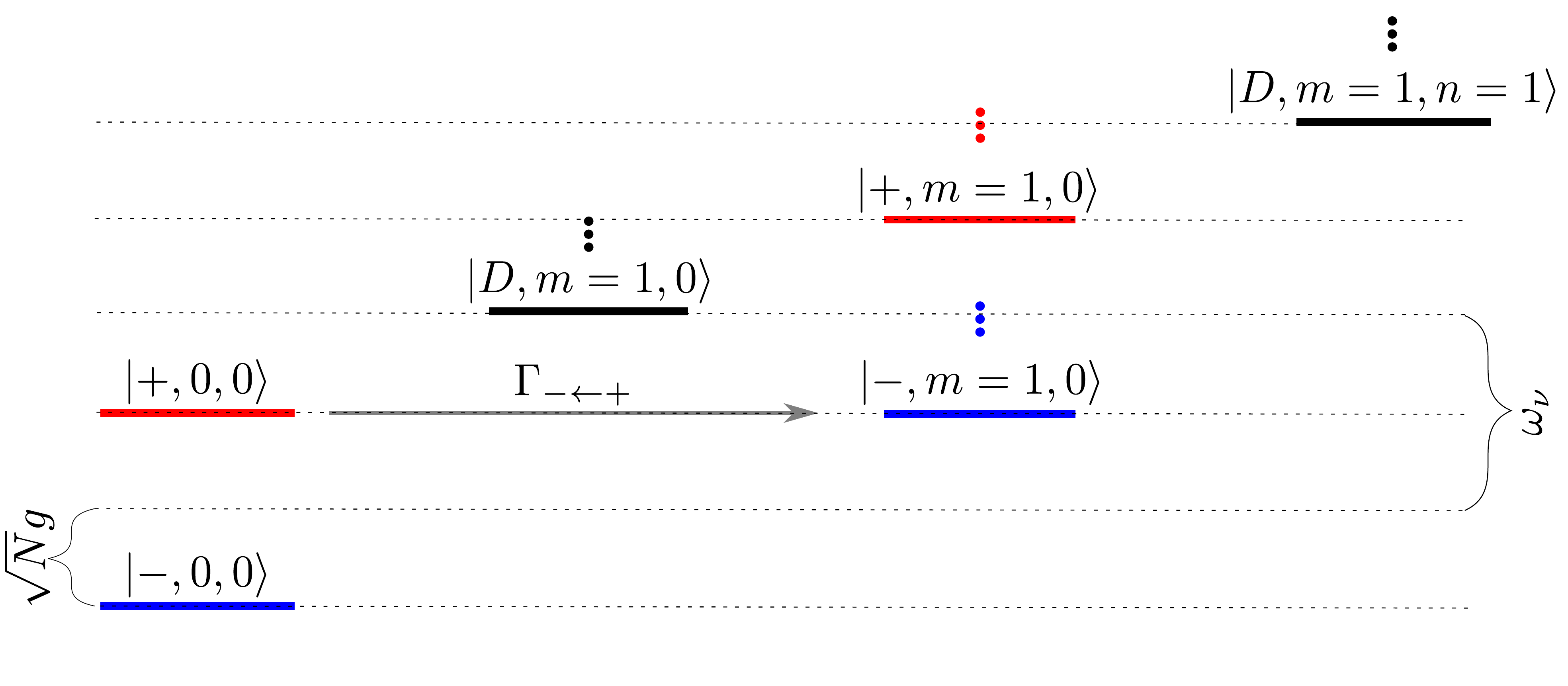}
\caption{Direct relaxation from the upper polariton to the lower polariton as described until the first-order correction (blue: lower polariton, red: upper polariton, black: dark states). States $|+,0,0\rangle$ and $|-,m=1,0\rangle$ are resonantly coupled through vibronic coupling to  modes of frequency $\omega_{\nu}=2\sqrt{N}g$.}
\label{decay2}
\end{center}
\end{figure*}

{\small

\begin{align}
\Gamma_{- \leftarrow +}&=2\pi\sum_{m}|\langle-,0,0|\hat{\tilde{H}}^{(1)}_{1}|+,0,0\rangle|^{2}\delta\left(\omega_{-,0,0}-\omega_{+,0,0}\right)\nonumber\\
&=\frac{2\pi}{4N}\sum_{k}\omega_{\nu,k}^{2}s_{k}\delta(2\sqrt{N}g-\omega_{\nu,k}).
\end{align}} Similarly, we can recalculate the decay rate from the upper polariton into the dark states, 

\begin{align}
&\Gamma_{D \leftarrow +}=2\pi\sum_{m}|\langle D,m,0|\hat{\tilde{H}}^{(1)}_{1}|+,0,0\rangle|^{2}\delta(\omega_{+,0,0}-\omega_{D,m,0})\nonumber\\
&=\left(\frac{N-1}{N}\right)\pi\sum_{k}\omega_{\nu,k}^{2}s_{k}\delta(\sqrt{N}g-\omega_{\nu,k}).
\end{align}where we have recovered the $\frac{N-1}{N}$ factor missing from Eq. \ref{eq:uptods1} in the zero-order approximation. The final state in the previous calculation can be used as initial state to describe the subsequent relaxation from the dark-states into the lower polariton,
{\small
\begin{align}
&\Gamma_{- \leftarrow D}=2\pi\sum_{m'}|\langle -,m',0|\hat{\tilde{H}}^{(1)}_{1}|-,0,0\rangle|^{2}\delta(\omega_{-,0,0}-\omega_{D,m=1,0})\nonumber\\
&=\left(\frac{N-1}{N^{2}}\right)\pi\sum_{k}\omega_{\nu,k}^{2}s_{k}\delta(\sqrt{N}g-\omega_{\nu,k}),
\end{align}}where the factor $\frac{N-1}{N^{2}}$ can be approximated as $1/N$. We suspect this factor will become exactly $1/N$ if higher order corrections are considered. The schematic representation of relaxation mechanisms is shown in Figs. \ref{decay1} and \ref{decay2}.

A summary of the results of this section is given in the next table.
\begingroup
\begin{longtable*}{|c||l||r|}
\hline
     & zeroth-order approximation& with first-order correction $\ \ \ \ $\\ 
\hline
    $\Gamma_{- \leftarrow +}$ & $\ \ \ \ \ \ \ \  \ \ \ \ \ \ \ 0$ & $\left(\frac{1}{4N}\right)2\pi\sum_{k}\omega_{\nu,k}^{2}s_{k}\delta(2\sqrt{N}g-\omega_{\nu,k})$\\ 
\hline
    $\Gamma_{D \leftarrow +}$ & $\pi\sum_{k}\omega_{\nu,k}^{2}s_{k}\delta(\sqrt{N}g-\omega_{\nu,k})$ & $\left(\frac{N-1}{N}\right)\pi\sum_{k}\omega_{\nu,k}^{2}s_{k}\delta(\sqrt{N}g-\omega_{\nu,k})$\\ 
\hline
   $\Gamma_{- \leftarrow D}$ & $\ \ \ \ \ \ \ \  \ \ \ \ \ \ \ 0$ & $\left(\frac{N-1}{N^{2}}\right)\pi\sum_{k}\omega_{\nu,k}^{2}s_{k}\delta(\sqrt{N}g-\omega_{\nu,k})$\\ 
\hline
\end{longtable*}
\endgroup
We conclude once more that in order to get dynamical effects that do not vanish in the thermodynamic limit it is sufficient to use the zeroth-order approximation. Likewise, to obtain slow rates proportional to $1/N$ (e.g., $\Gamma_{-\leftarrow +}$, $\Gamma_{-\leftarrow D}$) we only need to add the first-order correction. It is unlikely that we would need to go to higher-order corrections for microcavity polaritons, or even for polaritons arising in plasmonic antennas, where $N=100-1000$ \cite{Munkhbat}, although if strong coupling can be demonstrated with smaller $N$ values, those corrections could start mattering. On the other hand, it would be of interest to compare the performance of our method with the standard one of explicitly simulating several molecules using an optimized algorithm such as multilayer MCTDH \cite{Vendrellayer}, and characterize the values of $N$ after which our method can outcompete the latter.

\subsection{Finite Temperature effects}

In the previous sections we have only dealt with transitions from higher to lower lying polaritonic or dark states. To calculate rates such as $\Gamma_{D\leftarrow -}$ we need to consider all possible initial states allowed by thermal fluctuations. In general, such states involve breaking of the symmetry that is essential for the reduction of the dimensionality in Eqs. \ref{eq:EoM4}. Therefore, at the current stage, this formalism cannot be easily generalized to finite temperatures. Future works will focus on developing a density matrix approach in which permutational symmetries can be smoothly applied.	

\section{Non-statistical Excited-State Dynamics}\label{sec:nonstat}

In this section, we look at the mechanism whereby, given two molecular species strongly coupled to a cavity, excitation energy can be selectively funneled to one of the species. We will show that statistical yield estimates based on linear optical spectroscopy are misleading at predicting these outcomes accurately.

For a system with $N_\text{A}$ molecules of species $A$ and $N_\text{B}$ molecules of species $B$ inside a cavity, the Hamiltonian in the zeroth-order approximation (see Eq. \ref{eq:Makri}) can be readily generalized to,

\begin{widetext}
\begin{equation}
\hat{\tilde{H}}_0=
\left(\begin{array}{c|c c c|c c c}
 \omega_{c} & g_A\sqrt{N_A}\langle \varphi^{(A)}_{1}|\phi^{(A)}_{1}\rangle & \cdots & g_A\sqrt{N_B}\langle \varphi^{(A}_{1}|\phi^{(A)}_{m_A}\rangle  & g_A\sqrt{N_A}\langle \varphi^{(B)}_{1}|\phi^{(B)}_{1}\rangle & \cdots & g_B\sqrt{N_B}\langle \varphi^{(B)}_{1}|\phi^{(B)}_{m_B}\rangle \\
\hline
g_A\sqrt{N_A}\langle \varphi^{(A)}_{1}|\phi^{(A)}_{1}\rangle & \omega_{eg,1}^{(A)} & \cdots & 0  & 0 &  \cdots & 0\\

\vdots & 0  & \ddots & \vdots & \vdots & \ddots & \vdots\\

g_A\sqrt{N_A}\langle \varphi^{(A)}_{1}|\phi^{(A)}_{m_1}\rangle & \vdots & \hdots & \omega_{eg,m_A}^{(A)} & 0 & \hdots & 0\\
\hline
g_B\sqrt{N_B}\langle \varphi^{(B)}_{1}|\phi^{(B)}_{1}\rangle & 0 & \cdots & 0  & \omega_{eg,1}^{(B)} & \cdots  &  0\\

\vdots & \vdots & \ddots & \vdots  & 0  & \ddots & \vdots \\

g_B\sqrt{N_B}\langle \varphi^{(B)}_{1}|\phi^{(B)}_{m_B}\rangle & 0 & \cdots & 0  & 
\vdots & \hdots & \omega_{eg,m_B}^{(B)}
\end{array}\right)
\label{h0_lh}
\end{equation}
\end{widetext}

\begin{figure*}[htbp]
\begin{center}
\includegraphics[width=1\linewidth]{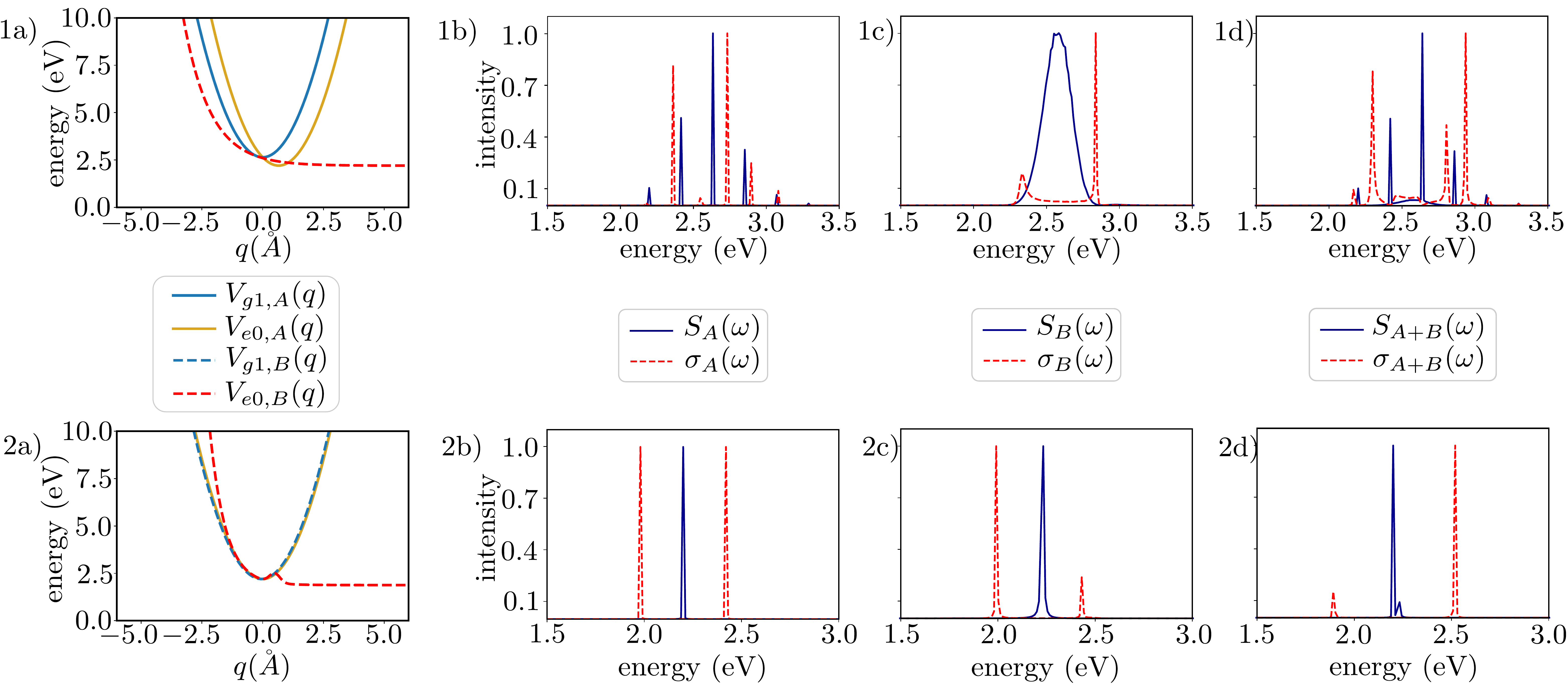}
\caption{Summary of molecular features used in Sec. \ref{sec:nonstat}, Examples 1 and 2. The relevant Potential Energy Surfaces (PESs) (1a,2a); horizontal axis should read $q=q_i$, depending on whether the PES refers to $i=A$ or $B$.  Linear absorption spectra outside (blue, $S(\omega)$) and inside (red, $\sigma(\omega)$) of cavity, corresponding to a ($1$b,$2$b) pure $A$, ($1$c,$2$c) pure $B$, and ($1$d,$2$d) $A/B$ mixture at $1:1$ ratio. In both examples, $A$ has a bound excited PES, while $B$ has a dissociative one. This has an obvious effect in the outside-of-cavity absorption for Example 1 ($1$b,$1$c) but not for Example 2 ($2$b,$2$c). The underlying reason for this discrepancy is that the dissociative character of $B$ is manifestly at the FC region in Example 1 ($1$a), while it is ``hidden" from the FC region in Example 2 ($2$a).}
\label{spec_pot}
\end{center}
\end{figure*}

\textit{Example 1.-} The molecular PESs in mass-weighted coordinates ($\mu=1$) are $ V_{g0,j}(q_{j}) = \frac{1}{2} \omega_g^2 q_{j}^{2}$ where $j\in \{A,B\}$ denote the molecular species, $V_{e0,A}(q_{A}) =  \frac{1}{2} \omega_{e}^2 (q_{A} - d_A)^{2} + \omega_{eg,1}^{(A)}$, and $V_{e0,B}(q_B) = e^{-a (q_{B} - d_B)}  + \omega_{eg,1}^{(B)}$. The excited PES of species $A$ is a displaced harmonic oscillator and that of species $B$ is a dissociative potential. For our simulations, we have chosen $\omega_g = \omega_e = 0.22$ eV, $d_A = d_B = 0.66$ \AA, $a=0.90$ \AA$^{-1}$, $\omega_{eg,1}^{(A)}=\omega_{eg,1}^{(B)}=2.2$ eV, and $g_A\sqrt{N_A}= g_B \sqrt{N_B}=0.22$ eV. The photon energy ($\omega_c$) is resonant with the FC transition of both $A$ and $B$ with $\omega_c = \omega_{eg,1}^{(A)} + \frac{1}{2} \omega_e^2 d_A^2$ (see Fig. \ref{spec_pot}, $1$a). \\

Given the one-dimensional nature of the PESs above, it is straightforward to explicitly construct $\hat{\tilde{H}}_0$ by computing the eigenstates and eigenenergies of the vibrational Hamiltonians; we use the standard Discrete Variable Representation method by Colbert and Miller \cite{Colbert}. This calculation was then used to compute spectra of the molecules outside and in the cavity using the formalism illustrated in \cite{Keeling, KeelingZeb}. The parameters for calculation of the spectra are chosen such that the frequency resolution and total simulation time satisfy Fourier transform relations. The total simulation time of $2.5$ ps determines the molecular and cavity linewidths as $0.002$ eV (see Fig. \ref{spec_pot}, $1$b,c,d). The absorption spectra of the bare molecules reveal the energy level structure accessed at the respective FC regions \cite{Heller}. For molecules $A$, we see the strongest peak corresponding to the FC transition accompanied by the other peaks of the vibronic progression. On the other hand, for molecules $B$, we see a single broad feature due to the dissociative potential. When strongly coupled to a cavity, these peaks form a rich pattern of peak splittings which can be intuitively understood upon diagonalization of Eq. \ref{h0_lh} (for instance, for molecule $A$, six sharp resonances become seven due to coupling to the cavity).

\begin{figure*}[htbp]
\begin{center}
\includegraphics[width=1\linewidth]{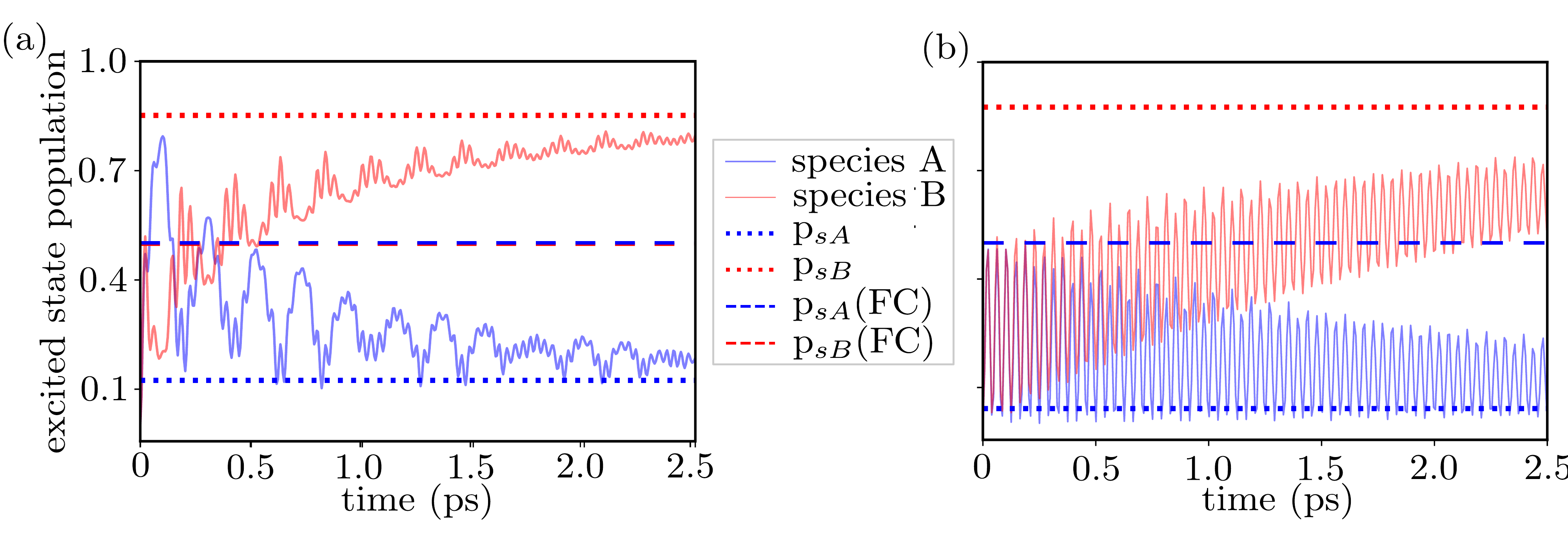}
\caption{Excited state populations of molecular species $A$ and $B$ as a function of time for (a) Example 1 and (b) Example 2. Dotted (dashed) horizontal lines show the estimated populations $p_{s_A}$, $p_{s_B}$ ($p_{s_A}$(FC),$ p_{s_B}$(FC)) from the eigenstates (``short-time eigenstates" defined at the FC point). Energy funneling from species $A$ into species $B$ is observed in both cases, indicating that this phenomenon is not easily predicted from linear optical spectra alone, since at least for Example 2, the absorption spectra of bare molecules $A$ and $B$ are pretty much identical (see Fig. \ref{spec_pot}, $2$b,c).} 
 \label{dynamics}
\end{center}
\end{figure*}

We build time-dependent wavefunctions by constructing the corresponding linear combinations of numerically computed eigenstates of $\hat{\tilde{H}}^{(0)}$. This procedure would obviously be impractical for realistic molecular species with many modes each, in which case, an explicit time-dependent approach such as MCTDH would be preferred. This simulation illustrates a phenomenon where the energy initially given to the cavity is eventually channeled preferentially to one of the two molecules. The results of this simulation are consistent with the phenomenology originally theoretically proposed by Groenhof and Toppari, based on computational simulations with at most 1000 molecules of one of the species \cite{Groenhof2}, and demonstrate the latter remains valid in the thermodynamic limit. For our simulation, we start with an excitation in the cavity, $|\Psi(0)\rangle=\varphi_{1}(q_{A})\varphi_{1}(q_{B})|1\rangle$. At short times the cavity evenly distributes the energy into both types of molecules, exciting them at their respective FC regions. We monitor the excited-state populations as a function of time (see Fig. \ref{dynamics}a). We observe that at short times, the population is transferred equally to both species. However, as time progresses, the excitation is funneled selectively to species $B$ via the cavity. The mechanism of the phenomenon is the following: at short times of the order of the Rabi oscillations, the cavity excites the bright modes of both species with equal populations at the respective FC regions due to the chosen equal collective light-matter couplings. However, as time progresses, excitations in species $B$ decay irreversibly into their dark states upon evolution away from the FC region (see Eq. \ref{eq:dspop}) due to the dissociative character of $V_{e0,B}(q_{B})$. Molecular species $A$ does not undergo this fast dephasing due to the bound nature of $V_{e0,A}(q_{A})$ and is the one that predominantly remains at its FC region and emits a photon to the cavity at the end of the Rabi cycle. The mechanism restarts with the re-excitation of both molecules by the photon. What results from this simulation is a net energy flow from molecular species $A$ to species $B$ mediated by the cavity, and which cannot be explained by the contribution of each molecule to the polariton states defined at the FC region. Notice that if the cavity decay is faster than the molecular dephasing, the excitation can leak out before energy transfer from $A$ to $B$ ensues. This issue can be overcome by having the cavity mode under continuous pumping. 

Next, we compare the computed long-time populations accumulated in the excited states of molecules $A$ and $B$ to the infinite-time probabilities of molecules of the cavity exciting the different species $( p_{s_j} \text{ for } j\in\{A,B\})$,
\begin{eqnarray}
    p_{s_j} &=&  \sum_n |\langle n | e_j\rangle|^2 |\langle n | 1\rangle|^2,
\label{stat}
\end{eqnarray}
Here, $\{|n\rangle\}$ represent the eigenstates of $\hat{\tilde{H}}^{(0)}$. Eq. \ref{stat} computes the composition of probabilities of an eigenstate simultaneously having photonic and molecule A(B) contribution. However, the entire set of eigenstates is not obtainable using linear absorption spectroscopy alone, given that the latter is a limited projection of information at the FC region \cite{Heller}. As an extreme example, we consider ``short-time-resolution eigenstates" consisting of the upper, middle, and lower polaritons computed at the FC region, and compute the analogous estimates $p_{s_A}$(FC) and $p_{s_B}$(FC). The infinite and short-time estimates are presented in Fig. \ref{dynamics}a. As expected, the plot suggests that the statistical estimate obtained from the full set of eigenvectors gives the right prediction of the yields at long times, while the FC estimate predicts the incorrect outcome of equal populations of $A$ and $B$. 

\textit{Example 2.-} Example 1 highlights the danger of inferring excited-state dynamics relying solely on information obtained from linear optical spectroscopy. This issue becomes even more pertinent owing to some observations of Xiang \emph{et. al.} \cite{Wei}, where the authors observe selective cavity-mediated energy transfer into one of two molecules, despite the bare linear absorption spectral lineshapes being very similar outside the cavity. To highlight some of the described subtleties, we explore a second example where we consider a slightly different shape of the excited state PES for species $B$ such that it resembles that of molecular species $A$ in the FC region, while still being dissociative. We now have $ V_{e0,B}(q_B) =e^{-a (q_{B} - d_{B,1})} + c e^{-b (q_{B} - d_{B,2})^2}+\omega_{eg,1}^{(B)}$ as the excited state potential of molecular species B, which is a dissociative potential with a bump. Otherwise, the parameters are $\omega_g=\omega_e=0.22$ eV, $d_A =0$, $a=1.54$ \AA$^{-1}$, $d_{B,1}=0.16$ \AA, $b=11.02$ \AA$^{-2}$, $c=0.51$ eV and $d_{B,2}=0.50$ \AA, and $\omega_{eg,1}^{(B)} = 1.88$ eV. The PESs for this case are shown in Fig. \ref{spec_pot}, $2$a.

The computed spectra for the bare species and their corresponding spectra under strong coupling to the cavity mode are shown in Fig. \ref{spec_pot}, $2$b,c,d. The resemblance of the excited state PESs at the FC region translates into very similar spectral lineshapes despite the dissociative \textit{vs} bound nature away from the FC region. Regardless of the similarities in the absorption spectra, starting the dynamics with an excitation in the cavity mode still gives rise to the effective energy transfer from species $A$ to $B$. Accordingly, the statistical ratio computed from the eigenstates at the FC point predicts equal populations of molecular species, while the full set of vibro-polaritonic eigenstates of the system ($p_{s_A}$ and $p_{s_B}$ in Fig.\ref{dynamics}b) predicts the correct yields. Since such states are not easily accessible from linear absorption spectroscopy, we consider those measurements insufficient to predict the outcome of these photoproducts.

It is important to note that by virtue of our reliance on the zeroth-order approximation, the cavity-mediated energy transfer mechanism studied in this section is $N$-independent, and thus differs from that put forward in Refs. \cite{MattPARET,Feist4}. The latter relies on a bottleneck transfer of population from dark to polariton modes, whose rate scales as $1/N$. 

In the future, it will be of great interest to ascertain whether the $\mathcal{O}(N^0)$ rate mechanism studied in the present work [originally put forward in \cite{Groenhof2}], or the mentioned $\mathcal{O}(N^{-1})$ rate mechanism is in order in the various experiments of cavity-mediated energy transfer \cite{Coles,Zhong,Wei}. At least, for the last reference, it is clear that the latter mechanism cannot be operative, for it would yield rates of energy transfer that are much slower than those observed in the experiment. 

Finally, we speculate that the mechanism highlighted in this section, where the cavity mediates energy transfer into the fastest dephasing molecular species, should be quite a generic phenomenon, and might be at play in the recent study in \cite{Giebink}, which shows polariton enhanced photoconductivity of an organic film. However, a further consideration of this problem merits a careful inclusion of disorder and temperature effects, and thus, is beyond the scope of this work.

\section{Summary}\label{sec:summ}

The method of Collective dynamics Using Truncated Equations (CUT-E) exploits permutational symmetries of an ensemble of identical molecules and an emergent hierarchy of timescales, to elucidate the excited-state dynamics under collective strong light-matter interaction. Although previous works have used equivalent symmetry arguments and have arrived at conclusions that are consistent with ours \cite{Keeling,KeelingZeb,Spano,Nitzan,Schafer}, elements of the present work uniquely provide opportunities for a systematic and intuitive treatment of molecular polariton dynamics. 

It is important to note that, via quite a different formalism, another method that maps the dynamics of molecular polaritons to a single effective molecule in a cavity has already recently been reported before our work \cite{Keeling}. This method, based on density matrices, does not provide $\mathcal{O}(1/\sqrt{N})$ corrections to dynamics, and considers a coherent state of the photon at all times, instead of the single-excitation manifold dynamics we have presented. We believe this method is quite complementary in its scope to ours. However, given the different formalisms, it is at present hard to assess the deeper conceptual connections between the methods; this will be subject of future work.

Let us conclude by highlighting some of the features of our approach. First, it allows for the systematic introduction of corrections to the thermodynamic limit. Second, our time-dependent approach generalizes some of the results found in previous work that also exploit permutational symmetries to compute optical properties \cite{Spano3,Spano2,Spano,Keeling,KeelingZeb,KeelingZeb2}; here we have generalized these concepts to chemical dynamics. Moreover, our work naturally provides an alternative interpretation of bright and dark-states based on permutationally-symmetric states, that is different from the one inherited from restrictive quantum optics models. These observations provide much needed physical intuition to design principles of polariton chemistry control, where rather than avoiding the decay into dark states, one embraces such phenomenon at one's advantage, such as with the examples provided in section \ref{sec:nonstat}. Finally, the method enjoys numerical simplicity and is written in a language that makes it straightforward to modify existing codes for single-molecule strong light-matter calculations, and more generally, convenient for implementation in existing quantum molecular dynamics algorithms.

\section{Acknowledgments}

This work was supported as part of the Center for Molecular Quantum Transduction (CMQT), an Energy Frontier Research Center funded by the U.S. Department of Energy, Office of Science, Basic Energy Sciences under Award No. DE-SC0021314. We also thank Kai Schwennicke and Matthew Du for useful discussions.

\bibliography{main}

\end{document}